%% file: hatp13.tex
\documentclass[apjl]{emulateapj}
\usepackage{comment}
\usepackage{ifthen}

\newcommand{\ctbd}[1]{}

\newcommand{\lc}{light curve}
\newcommand{\lcs}{light curves}
\newcommand{\Lc}{Light curve}

\newcommand{\band}[1]{\ensuremath{#1}~band}

\newcommand{\chisq}{\ensuremath{\chi^2}}

\newcommand{\kms}{\ensuremath{\rm km\,s^{-1}}}
\newcommand{\ms}{\ensuremath{\rm m\,s^{-1}}}

\newcommand{\gcmc}{\ensuremath{\rm g\,cm^{-3}}}
\newcommand{\ergscmsq}{\ensuremath{\rm erg\,s^{-1}\,cm^{-2}}}

\newcommand{\vsini}{\ensuremath{v \sin{i}}}
\newcommand{\feh}{\ensuremath{\rm [Fe/H]}}

\newcommand{\vmac}{\ensuremath{v_{\rm mac}}}
\newcommand{\vmic}{\ensuremath{v_{\rm mic}}}
\newcommand{\rhk}{\ensuremath{R^{\prime}_{HK}}}
\newcommand{\logrhk}{\ensuremath{\log\rhk}}

\newcommand{\rsun}{\ensuremath{R_\sun}}
\newcommand{\msun}{\ensuremath{M_\sun}}
\newcommand{\lsun}{\ensuremath{L_\sun}}

\newcommand{\rstar}{\ensuremath{R_\star}}
\newcommand{\mstar}{\ensuremath{M_\star}}
\newcommand{\lstar}{\ensuremath{L_\star}}

\newcommand{\teffstar}{\ensuremath{T_{\rm eff\star}}}
\newcommand{\rhostar}{\ensuremath{\rho_\star}}
\newcommand{\loggstar}{\ensuremath{\log{g_{\star}}}}

\newcommand{\mearth}{\ensuremath{M_\earth}}

\newcommand{\rpl}{\ensuremath{R_{p}}}
\newcommand{\mpl}{\ensuremath{M_{p}}}

\newcommand{\rhopl}{\ensuremath{\rho_{p}}}
\newcommand{\ipl}{\ensuremath{i_{p}}}

\newcommand{\arstar}{\ensuremath{a/\rstar}}
\newcommand{\zrstar}{\ensuremath{\zeta/\rstar}}

\newcommand{\rjup}{\ensuremath{R_{\rm J}}}
\newcommand{\mjup}{\ensuremath{M_{\rm J}}}

\newcommand{\reffig}[1]{Fig.~\ref{fig:#1}}
\newcommand{\refsec}[1]{\mbox{\S\ \ref{sec:#1}}}

\newcommand{\reftab}[1]{Tab.~\ref{tab:#1}}

\newcommand{\flwof}{\mbox{FLWO 1.2\,m}}

\newcommand{\flwos}{\mbox{FLWO 1.5\,m}}

\newcommand{\hd}[1]{\mbox{HD #1}}

\newcommand{\tbn}[1]{\tablenotemark{#1}}

\input{planetinfo.tex}

\newcommand{\hatcur}{HAT-P-13}
\newcommand{\hatcurb}{HAT-P-13b}
\newcommand{\hatcurc}{HAT-P-13c}
\newcommand{\hatcurRVgammarel}{\hatcurRVgamma}                           
\newcommand{\hatcurSMEversion}{ii}                                        
\newcommand{\hatcurSMEteff}{\ifthenelse{\equal{\hatcurSMEversion}{i}}{\hatcurSMEiteff}{\hatcurSMEiiteff}}
\newcommand{\hatcurSMEzfeh}{\ifthenelse{\equal{\hatcurSMEversion}{i}}{\hatcurSMEizfeh}{\hatcurSMEiizfeh}}
\newcommand{\hatcurSMEzfehshort}{\ifthenelse{\equal{\hatcurSMEversion}{i}}{\hatcurSMEizfehshort}{\hatcurSMEiizfehshort}}
\newcommand{\hatcurSMElogg}{\ifthenelse{\equal{\hatcurSMEversion}{i}}{\hatcurSMEilogg}{\hatcurSMEiilogg}}
\newcommand{\hatcurSMEvsin}{\ifthenelse{\equal{\hatcurSMEversion}{i}}{\hatcurSMEivsin}{\hatcurSMEiivsin}}
\newcommand{\hatcurSMEvmac}{\ifthenelse{\equal{\hatcurSMEversion}{i}}{\hatcurSMEivmac}{\hatcurSMEiivmac}}
\newcommand{\hatcurSMEvmic}{\ifthenelse{\equal{\hatcurSMEversion}{i}}{\hatcurSMEivmic}{\hatcurSMEiivmic}}

\shortauthors{Bakos et al.}
\shorttitle{\hatcur\lowercase{b,c}}

\newboolean{emulateapj}
\setboolean{emulateapj}{true}

\newboolean{rvtablelong}
\setboolean{rvtablelong}{true}

\ifthenelse{\boolean{emulateapj}}{
	\newcommand{\titledag}{$\dagger$}
}{
	\newcommand{\titledag}{\dagger}
}

\begin{document}

\title{\hatcur\lowercase{b,c}: a transiting hot Jupiter with a massive
	outer companion on an eccentric orbit\altaffilmark{\titledag}}

\author{
	G.~\'A.~Bakos\altaffilmark{1,2},
	A.~W.~Howard\altaffilmark{3},
	R.~W.~Noyes\altaffilmark{1},
	J.~Hartman\altaffilmark{1},
	G.~Torres\altaffilmark{1},
	G\'eza~Kov\'acs\altaffilmark{4},
	D.~A.~Fischer\altaffilmark{5},
	D.~W.~Latham\altaffilmark{1},
	J.~A.~Johnson\altaffilmark{6},
	G.~W.~Marcy\altaffilmark{3},
	D.~D.~Sasselov\altaffilmark{1},
	R.~P.~Stefanik\altaffilmark{1},
	B.~Sip\H{o}cz\altaffilmark{1,7},
	G\'abor~Kov\'acs\altaffilmark{1},
	G.~A.~Esquerdo\altaffilmark{1},
	A.~P\'al\altaffilmark{4,1},
	J.~L\'az\'ar\altaffilmark{8},
	I.~Papp\altaffilmark{8},
	P.~S\'ari\altaffilmark{8}
}
\altaffiltext{1}{Harvard-Smithsonian Center for Astrophysics,
	Cambridge, MA, gbakos@cfa.harvard.edu}

\altaffiltext{2}{NSF Fellow}

\altaffiltext{3}{Department of Astronomy, University of California,
	Berkeley, CA}

\altaffiltext{4}{Konkoly Observatory, Budapest, Hungary}

\altaffiltext{5}{Department of Physics and Astronomy, San Francisco
	State University, San Francisco, CA}

\altaffiltext{6}{Institute for Astronomy, University of Hawaii,
Honolulu, HI 96822; NSF Postdoctoral Fellow}

\altaffiltext{7}{Department of Astronomy,
	E\"otv\"os Lor\'and University, Budapest, Hungary.}

\altaffiltext{8}{Hungarian Astronomical Association, Budapest, 
	Hungary}

\altaffiltext{$\dagger$}{
	Based in part on observations obtained at the W.~M.~Keck
	Observatory, which is operated by the University of California and
	the California Institute of Technology. Keck time has been
	granted by NOAO (A146Hr,A264Hr) and NASA (N128Hr,N145Hr).
}

\begin{abstract}

We report on the discovery of a planetary system with a close-in
transiting hot Jupiter on a near circular orbit and a massive outer planet
on a highly eccentric orbit. The inner planet, \hatcurb, transits the
bright V=\hatcurCCtassmv\ \hatcurISOspec\ dwarf star \hatcurCCgsc\
every $P=\hatcurLCP$~days, with transit epoch $T_c = \hatcurLCT$ (BJD)
and duration \hatcurLCdur\,d. The outer planet, \hatcurc\ orbits the
star with $P_2 = \hatcurcLCP$~days and nominal transit center (assuming
zero impact parameter) of $T_{2c} = \hatcurcLCT$ (BJD) or time of
periastron passage $T_{2,peri}=\hatcurcPPperi$ (BJD). Transits of the
outer planet have not been observed, and may not be present. The host
star has a mass of \hatcurISOm\,\msun, radius of \hatcurISOr\,\rsun,
effective temperature \hatcurSMEteff\,K, and is rather metal rich with
$\feh = \hatcurSMEzfeh$. The inner planetary companion has a mass of
\hatcurPPmlong\,\mjup, and radius of \hatcurPPrlong\,\rjup\ yielding a
mean density of \hatcurPPrho\,\gcmc. The outer companion has $m_2\sin
i_2 = \hatcurcPPmlong\,\mjup$, and orbits on a highly eccentric orbit
of $e_2 = \hatcurcRVeccen$.
While we have not detected significant transit timing variations of
\hatcurb, due to gravitational and light-travel time effects, future
observations will constrain the orbital inclination of \hatcurc, along
with its mutual inclination to \hatcurb. 
The \hatcur\,(b,c) double-planet system may
prove extremely valuable for theoretical studies of the formation and 
dynamics of planetary systems.

\end{abstract}



\section{Introduction}
\label{sec:introduction}

Radial velocity (RV) surveys have shown that multiple-planet stellar
systems are common. For example, \citet{wright:2007} concluded that the
occurrence of additional planets among stars already having one known
planet must be greater than 30\%.  Thus, one would expect that out of
the $\sim50$ published transiting extrasolar planet (TEP)
systems\footnote{www.exoplanet.eu}, there should be a number of systems
with additional companions; these companions should make their presence
known through radial velocity variations of the parent stars, even if
they do not themselves transit.  The fact that so far no TEPs in
multiple planet systems have been reported is somewhat surprising based
on the statistics from RV surveys \citep[see also][]{fabrycky:2009}.
Recently \citet{smith:2009} searched the \lcs\ of 24 transiting planets
for outer companions, finding no evidence for a double transiting
system.

The Hungarian-made Automated Telescope Network (HATNet) survey
\citep{bakos:2002,bakos:2004} has been a major contributor to the
discovery of TEPs. Operational since 2003, it has covered approximately
10\% of the Northern sky, searching for TEPs around bright stars
($8\lesssim I \lesssim 12.5$\,mag). HATNet operates six wide field
instruments: four at the Fred Lawrence Whipple Observatory (FLWO) in
Arizona, and two on the roof of the Submillimeter Array hangar (SMA) of
SAO at Hawaii. Since 2006, HATNet has announced and published 12 TEPs.
A study similar to that of \citet{smith:2009} has been carried out on
9 known transiting planets from the HATNet project by Fabrycky (private
communication) with a similar null result.

In this work we report on the 13th discovery of HATNet, the detection
of the first known system with a transiting inner planet (\hatcurb)
which also contains a second, outer planet (\hatcurc), as detected by
the radial velocity variation of the host star. There have been
examples of transiting systems where the RV variations do show a long
term trend, such as HAT-P-7b \citep{winn:2009} and 
HAT-P-11b \citep{bakos:2009}, but no orbital
solution has yet been presented for any of these outer companions,
simply because there has not been a long enough timespan to cover the
period, or at least observe the long term trend changing sign. 
While no  transits of \hatcurc\ have yet been detected, the 
probability that the
companion actually transits the star is non-negligible if the orbits of
the two planets are nearly co-planar (the pure geometric transit
probability for \hatcurc\ is 1.3\%, see \refsec{hatcurc}). The system is
particularly interesting because the outer planet has both a high
eccentricity and a very high mass. These properties, in turn, should
induce transit timing variations (TTVs) of the inner planet of the
order of 10 seconds \citep[standard deviation,][]{agol:2005}. Such TTVs
may be used to constrain the orbital parameters of the outer planet,
including the inclination with respect to the line of sight and with
respect to the orbital plane of the inner planet.

In \refsec{obs} of this paper we summarize the observations, including
the photometric detection, and follow-up observations.
\refsec{analysis} describes the analysis of the data, such as the
stellar parameter determination (\refsec{stelparam}), blend modeling
(\refsec{blend}), and global modeling of the data (\refsec{globmod}).
We discuss our findings in \refsec{discussion}.

\section{Observations}
\label{sec:obs}

\subsection{Photometric detection}
\label{sec:detection}

The transits of \hatcurb{} were detected with the HAT-5 telescope. The
region around \hatcurCCgsc{}, a field internally labeled as
\hatcurfield, was observed on a nightly basis between 2005 November 25
and 2006 May 20, whenever weather conditions permitted. We gathered
4021 exposures of 5 minutes at a 5.5-minute cadence. Each image
contained approximately 20000 stars down to $I\sim14.0$. For the
brightest stars in the field we achieved a per-image photometric
precision of 3.1\,mmag.

\begin{figure}[!ht]
\plotone{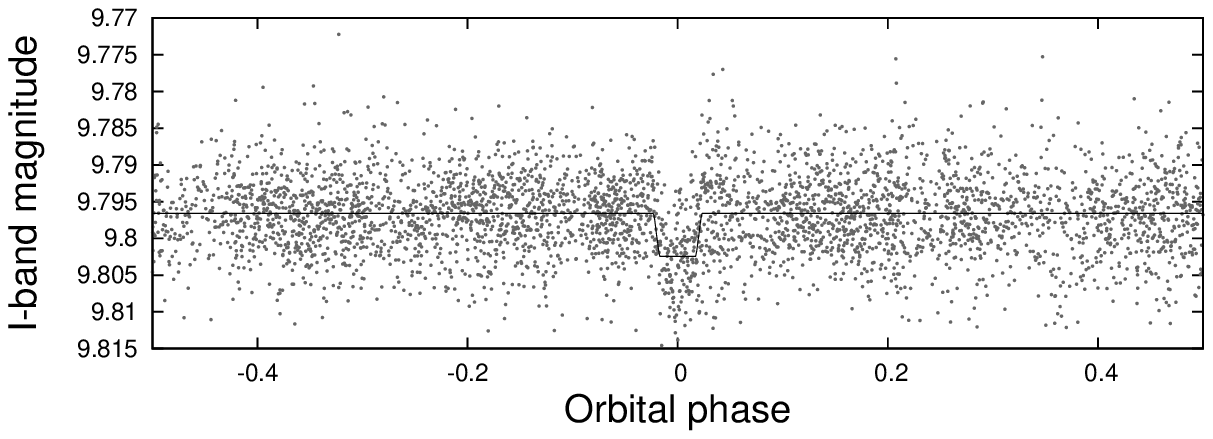}
\caption{
	The unbinned \lc{} of \hatcur{} including all 4021 instrumental
	\band{I} measurements obtained with the HAT-5 telescope
	of HATNet (see text for details), and folded with the period of $P
	= \hatcurLCPprec$\,days (which is the result of the fit described
	in \refsec{analysis}).
\label{fig:hatnet}}
\end{figure}

The calibration of the HATNet frames was done utilizing standard
procedures. The calibrated frames were then subjected to star detection
and astrometry, as described in \cite{pal:2006}. Aperture photometry
was performed on each image at the stellar centroids derived from the
2MASS catalog \citep{skrutskie:2006} and the individual astrometrical
solutions. Description of numerous details related to the reduction are
also given in \citet{pal:2009}. The resulting \lcs\ were decorrelated
against trends using the External Parameter Decorrelation technique in
``constant'' mode \citep[EPD, see][]{bakos:2009} and the Trend
Filtering Algorithm \citep[TFA, see][]{kovacs:2005}. The \lcs{} were
searched for periodic box-like signals using the Box Least Squares
method \citep[BLS, see][]{kovacs:2002}. We detected a significant 
signal in the \lc{} of \hatcurCCgsc{} (also known as
\hatcurCCtwomass{}; $\alpha = \hatcurCCra$, $\delta = \hatcurCCdec$;
J2000), with a depth of $\sim\hatcurLCdip$\,mmag, and a period of
$P=\hatcurLCPshort$\,days. The dip
significance parameter \citep[][]{kovacs:2002} of the signal was DSP =
17. The dip had a relative duration (first to
last contact) of $q\approx\hatcurLCq$, corresponding to a total
duration of $Pq\approx\hatcurLCdurhr$~hours (see \reffig{hatnet}).

\vspace{10mm}
\subsection{Reconnaissance Spectroscopy}
\label{sec:recspec}

One of the important tools used in our survey for establishing whether
the transit-feature in the \lc\ of a candidate is due to astrophysical
phenomena other than a planet transiting a star is the CfA Digital
Speedometer \citep[DS;][]{latham:1992}, mounted on the \flwos\
telescope. This yields high-resolution spectra with low signal-to-noise
ratio sufficient to derive radial velocities with moderate precision
(roughly 1\,\kms), and to determine the effective temperature and
surface gravity of the host star. With this facility we are able to
weed out certain false alarms, such as eclipsing binaries and multiple
stellar systems.

We obtained \hatcurDSnumspec\ spectra spanning \hatcurDSspan\ days with
the DS. The RV measurements of \hatcur{} showed an rms residual of
\hatcurDSrvrms\,\kms, consistent with no detectable RV variation. The
spectra were single-lined, showing no evidence for more than one star
in the system.
Atmospheric parameters for the star, including the initial estimates of
effective temperature $\teffstar = \hatcurDSteff\,K$, surface gravity
$\loggstar = \hatcurDSlogg$ (log cgs), and projected rotational
velocity $\vsini = \hatcurDSvsini\,\kms$, were derived as described by
\cite{torres:2002}. The effective temperature and surface gravity
correspond to a G4 dwarf \citep{cox:2000}. The mean line-of-sight
velocity of \hatcur\ is \hatcurDSgamma\,\kms.

\subsection{High resolution, high S/N spectroscopy and the search
for radial velocity signal components}
\label{sec:hispec}

\begin{figure}[!ht]
\plotone{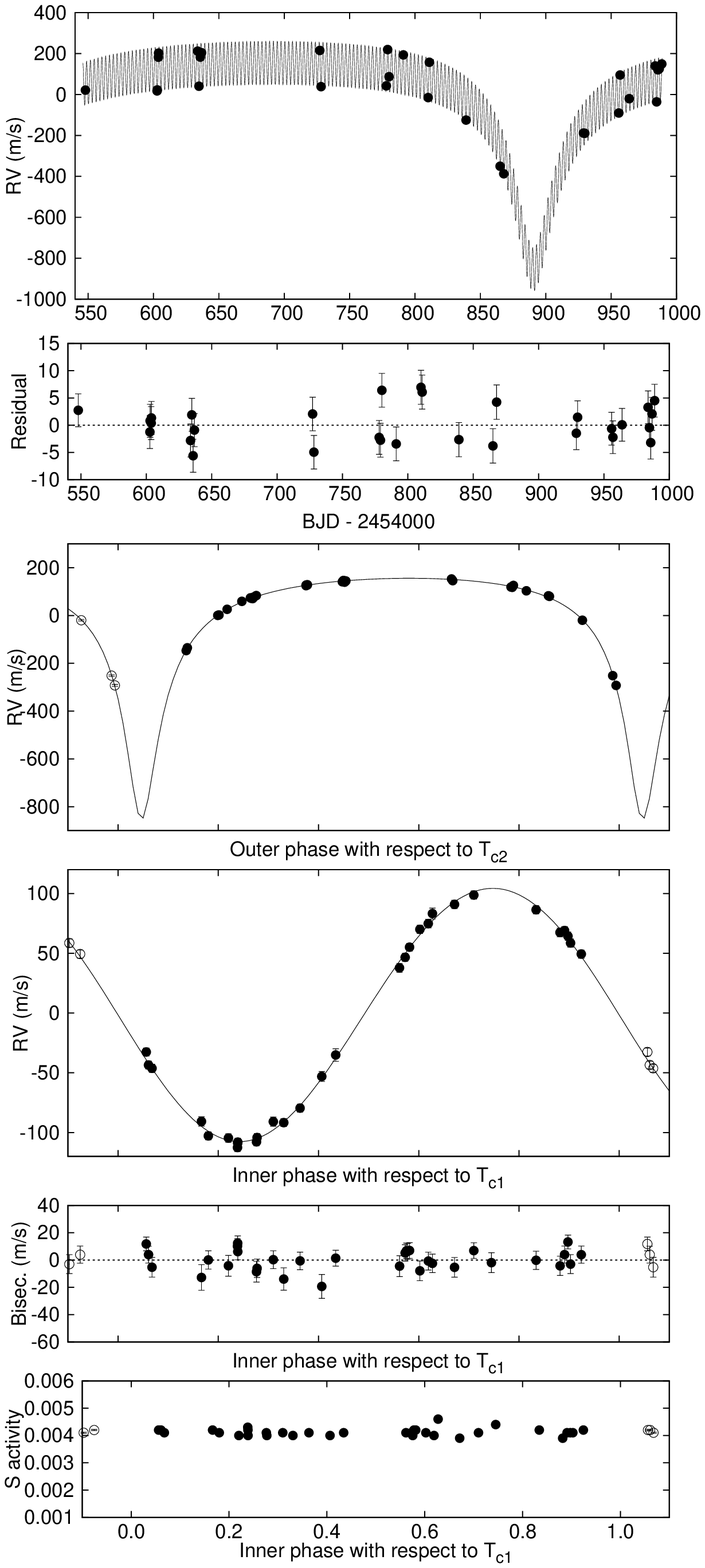}
\caption{
	(Top) Radial-velocity measurements from Keck for \hatcur{}, along
	with the best 2-planet orbital fit, shown as a function of BJD (see
	\refsec{analysis}).  The center-of-mass velocity has been
	subtracted.
	(Second panel) Phased residuals after subtracting the orbital fit
	(also see \refsec{analysis}). The rms variation of the residuals is
	about 3.4\,\ms, requiring a jitter of $\hatcurRVjitter$\,\ms\ to be
	added in quadrature to the individual errors to yield a reduced
	$\chisq$ of 1.0. The error-bars in this panel have been inflated
	accordingly. Note the different vertical scale of the panels.
	(Third panel) Orbit of the outer planet after subtracting the
	orbital fit of the inner planet from the data. Zero phase is
	defined by the fictitious transit midpoint of the second planet
	(denoted as $T_{c2}$, where $c$ is the common subscript for
	``center'', and ``2'' refers to the second planet). 
	The error-bars (inflated by the jitter) are smaller than the size
	of the points.
	(Fourth panel) Orbit of the inner planet after subtracting the
	orbital fit of the outer planet. Zero-phase is defined by the
	transit midpoint of \hatcurb\ (denoted as $T_{c1}$). 
	The error-bars are smaller than the 
	size of the points.
	(Fifth panel) Bisector spans (BS) for the Keck spectra including
	the two template spectra (\refsec{blend}).  The mean value has been
	subtracted.
	(Bottom) Relative S activity index values for the Keck spectra
	\citep[see][]{hartman:2009}. Open circles in the phase-plots
	indicate data that appears twice due to plotting outside the
	[0,1] phase domain. 
\label{fig:rv}}
\end{figure}

Given the significant transit detection by HATNet, and the encouraging
DS results, we proceeded with the follow-up of this candidate by
obtaining high-resolution and high S/N spectra to characterize the
radial velocity variations and to determine the stellar parameters with
higher precision. Using the HIRES instrument \citep{vogt:1994} on the
Keck~I telescope located on Mauna Kea, Hawaii, we obtained 30 exposures
with an iodine cell, plus three iodine-free templates. The first
template had low signal-to-noise ratio, thus we repeated the template
observations during a later run, and acquired two high quality
templates. We used the last two templates in the analysis. 
The observations were made between 2008 March 22 and 2009 June 5.
The relative radial velocity (RV) measurements are listed in 
\reftab{rvs}, and shown in \reffig{rv}.

\ifthenelse{\boolean{emulateapj}}{
    \begin{deluxetable*}{lrrrrrr}
}{
    \begin{deluxetable}{lrrrrrr}
}
\tablewidth{0pc}
\tablecaption{
	Relative radial velocity, bisector and activity index
	measurements of \hatcur{}.
	\label{tab:rvs}
}
\tablehead{
	\colhead{BJD} & 
	\colhead{RV\tablenotemark{a}} & 
	\colhead{\ensuremath{\sigma_{\rm RV}}\tablenotemark{b}} & 
	\colhead{BS} & 
	\colhead{\ensuremath{\sigma_{\rm BS}}} & 
	\colhead{S\tbn{c}} & 
	\colhead{\ensuremath{\sigma_{\rm S}}}\\
	\colhead{\hbox{~~~~(2,454,000$+$)~~~~}} & 
	\colhead{(\ms)} & 
	\colhead{(\ms)} &
	\colhead{(\ms)} &
    \colhead{(\ms)} &
	\colhead{} &
	\colhead{}
}
\startdata
\ifthenelse{\boolean{rvtablelong}}{
	\input{data/rvtable.tex}
}{
	\input{data/rvtable_short.tex}
}
\enddata
\tablenotetext{a}{
	The fitted zero-point that is on an arbitrary scale (denoted as
	$\gamma_{rel}$ in \reftab{planetparam}) has been
	subtracted from the velocities.
}
\tablenotetext{b}{
        The values for \ensuremath{\sigma_{\rm RV}} do not include
        the jitter.
}
\tablenotetext{c}{
		S values are on a relative scale. 
}
\ifthenelse{\boolean{rvtablelong}}{
}{
	\tablecomments{
		This table is presented in its entirety in the electronic edition
		of the Astrophysical Journal. A portion is shown here for guidance
		regarding its form and content.
}
}
\ifthenelse{\boolean{emulateapj}}{
    \end{deluxetable*}
}{
    \end{deluxetable}
}

Observations and reductions have been carried out in an identical way
to that described in earlier HATNet discovery papers, such as
\citet[][]{bakos:2009}. References for
the Keck iodine cell observations and the reduction of the radial
velocities are given in \citet{marcy:1992} and \citet{butler:1996}.

Initial fits of these RVs to a single-planet Keplerian orbit were quite
satisfactory but soon revealed a slight residual trend that became more
significant with time. This is the reason for the observations
extending over more than one year, as opposed to just a few months as
necessary to confirm simpler purely sinusoidal variations seen in other
transiting planets discovered by HATNet. Eventually we noticed that the
residual trend reversed (\reffig{rv}, top), a clear sign of a coherent
motion most likely due to a more distant body in the same system,
possibly a massive planet. Preliminary two-planet orbital solutions
provided a much improved fit (although with a weakly determined outer
period due to the short duration of the observations), and more
importantly, held up after additional observations. With the data so
far gathered and presented in this work, a false alarm probability
(FAP) analysis for the Keplerian orbit of the second planet yielded an
FAP of 0.00001.

A more thorough analysis using a two-component least squares Fourier
fit \citep[see][for the single-component case]{barning:1963}, with one
component fixed to the known frequency of the short-period inner planet,
also confirmed the existence of the long period component, and
indicated that the latter is due to a highly non-sinusoidal motion. A
number of other low frequency peaks were eliminated as being aliases
yielding worse quality fits. A three-harmonic representation of the
fit yielded an RMS of $5$\,\ms, fairly close to the value expected from
the formal errors. Full modeling of this complex motion of two
Keplerian orbits superposed, simultaneously with the photometry, is
described in our global fit of \refsec{globmod}.

\subsection{Photometric follow-up observations}
\label{sec:phot}

\begin{figure}[!ht]
\plotone{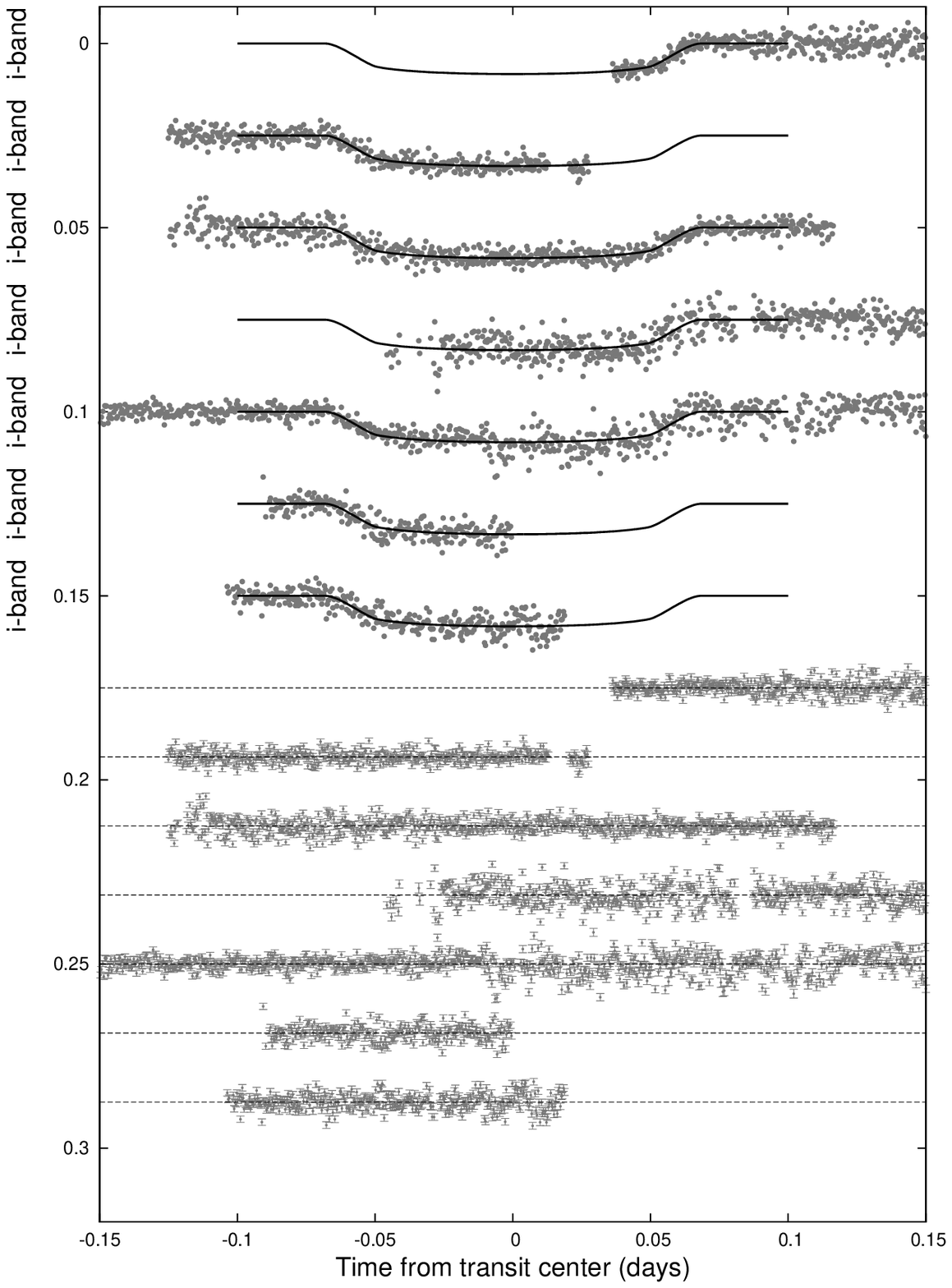}
\caption{
	Unbinned instrumental \band{i} transit \lcs{}, acquired with
	KeplerCam at the \flwof{} telescope on seven different dates. If
	the first full transit is assigned $N_{tr} = 0$ (2008 November 8/9
	MST, the third event from the top), then the follow-up \lcs\ have
	$N_{tr} = $ -68, -1, 0, 1, 24, 35, 62 from top to bottom.
	%
	%
	Superimposed are the best-fit transit model \lcs\ as described in
	\refsec{analysis}.  In the bottom of the figure we show the
	residuals from the fit.  Error-bars represent the photon and
	background shot-noise, plus the readout noise.
\label{fig:lc}}
\end{figure}

We observed 7 transit events of \hatcur{} with
the KeplerCam CCD on the \flwof{} telescope between 2008 April 24 and
2009 May 8. All observations were carried out in the \band{i}, and the
typical exposure time was 15 seconds.
%
%
The reduction of the images, including calibration, astrometry, and
photometry, was performed as described in \citet{bakos:2009}. 

We performed EPD and TFA against trends simultaneously with the \lc\
modeling (for more details, see \refsec{analysis}). From the series of
apertures, for each night, we chose the one yielding the smallest
residual rms for producing the final \lc.  These are shown in the upper
plot of \reffig{lc}, superimposed with our best fit transit \lc{}
models (see also \refsec{analysis}).

\begin{deluxetable}{lrrrr}
\tablewidth{0pc}
\tablecaption{Photometry follow-up of \hatcur\label{tab:phfu}}
\tablehead{
	\colhead{BJD} & 
	\colhead{Mag\tablenotemark{a}} & 
	\colhead{\ensuremath{\sigma_{\rm Mag}}} &
	\colhead{Mag(orig)} & 
	\colhead{Filter} \\
	\colhead{\hbox{~~~~(2,400,000$+$)~~~~}} & 
	\colhead{} & 
	\colhead{} &
	\colhead{} & 
	\colhead{}
}
\startdata
$ 54581.66038 $ & $   0.00550 $ & $   0.00080 $ & $   0.03620 $ &  i\\
$ 54581.66070 $ & $   0.00763 $ & $   0.00080 $ & $   0.03652 $ &  i\\
$ 54581.66104 $ & $   0.00877 $ & $   0.00080 $ & $   0.03686 $ &  i\\
$ 54581.66138 $ & $   0.00725 $ & $   0.00080 $ & $   0.03720 $ &  i\\
$ 54581.66171 $ & $   0.00670 $ & $   0.00080 $ & $   0.03753 $ &  i\\
$ 54581.66205 $ & $   0.01004 $ & $   0.00080 $ & $   0.03787 $ &  i\\
$ 54581.66237 $ & $   0.00673 $ & $   0.00080 $ & $   0.03819 $ &  i\\
$ 54581.66271 $ & $   0.00787 $ & $   0.00080 $ & $   0.03853 $ &  i\\
\enddata
\tablenotetext{a}{
	The out-of-transit level has been subtracted. These magnitudes have
	been subjected to the EPD and TFA procedures (in ``ELTG'' mode),
	carried out simultaneously with the transit fit.
}
\tablecomments{
	This table is presented in its entirety in the electronic edition
	of the Astrophysical Journal. A portion is shown here for guidance
	regarding its form and content.
}
\end{deluxetable}

\section{Analysis}
\label{sec:analysis}

\subsection{Properties of the parent star}
\label{sec:stelparam}

We derived the initial stellar atmospheric parameters by using
the first  template spectrum obtained with the Keck/HIRES instrument. We used
the SME package of \cite{valenti:1996} along with the atomic-line
database of \cite{valenti:2005}, which yielded the following {\em
initial} values and uncertainties (which we have conservatively
increased, to include our estimates of the systematic errors):
effective temperature $\teffstar=\hatcurSMEiteff$\,K, 
stellar surface gravity $\loggstar=\hatcurSMEilogg$\,(cgs),
metallicity $\feh=\hatcurSMEizfeh$\,dex, and 
projected rotational velocity $\vsini=\hatcurSMEivsin\,\kms$.

Further analysis of the spectra has shown that the host star is
chromospherically quiet with $\logrhk=-5.10$ and $S = 0.14$ (on an
absolute scale).

To determine the stellar properties via a set of isochrones, we used
three parameters: the stellar effective temperature, the metallicity,
and the normalized semi-major axis \arstar\ (or related mean stellar
density \rhostar). We note that another possible parameter in place of
\arstar\ would be the stellar surface gravity, but in the case of
planetary transits the \arstar\ quantity typically imposes a stronger
constraint on the stellar models \citep{sozzetti:2007}. (The validity
of this assumption, namely that the adequate physical model describing
our data is a planetary transit, as opposed to e.g.~a blend, is shown
later in \refsec{blend}.) 
%
%
With quadratic limb darkening coefficients (listed in \reftab{stellar})
from \cite{claret:2004} appropriate for the values of \teffstar, \feh,
and \loggstar as derived from the SME analysis, we performed a global
modeling of the data (\refsec{globmod}), yielding a full Monte-Carlo
distribution of \arstar. For \teffstar\ and \feh we adopted normal
distributions in the Monte Carlo analysis, with dispersions equal to
the 1-$\sigma$ uncertainties from the initial SME analysis.

For each
combination within the large ($\sim20000$) set of \arstar, \teffstar,
\feh\ values, we searched the stellar isochrones of the \citet{yi:2001}
models for the best fit stellar model parameters (\mstar, \rstar, age,
\loggstar, etc.).
We derived the mean values and uncertainties
of the physical parameters based on their {\em a posteriori}
distribution \citep[see e.g.][]{pal:2008}.

\begin{figure}[!ht]
\plotone{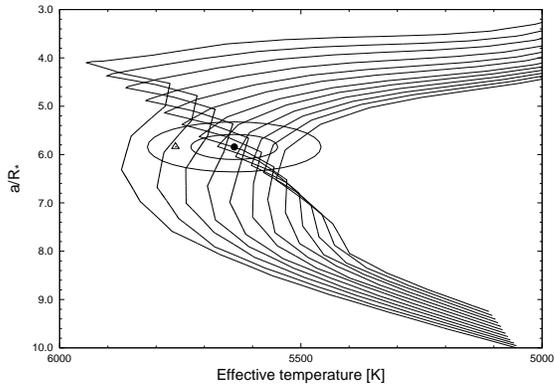}
\caption{
	Stellar isochrones from \citet{yi:2001} for metallicity
	\feh=\hatcurSMEiizfehshort\ and ages between 3.6 and 8.0\,Gyr in
	steps of 0.4\,Gyr. The final choice of $\teffstar$ and \arstar\ are
	marked and encircled by the 1-$\sigma$ and 2-$\sigma$ confidence
	ellipsoids. The open triangle denotes the $\teffstar$,\arstar\
	point found during the first SME iteration, before refining the
	stellar surface gravity. 
\label{fig:iso}}
\end{figure}

We then repeated the SME analysis by fixing the stellar surface gravity
to the refined value of $\loggstar=\hatcurISOlogg$ based on the
isochrone search, and only adjusting \teffstar, \feh\ and \vsini.  The
SME analysis was performed on the first, weaker template observation,
and also on the second and third, higher S/N pair of template
observations taken by Keck/HIRES. While the pair of high S/N templates
were acquired very close in time, and the respective SME values were
consistent to within a small fraction of the formal error-bars, they
were also consistent to within 1-$\sigma$ with the values based on the
weaker template that was taken much earlier. This consistency reassured
that our stellar atmospheric parameter determination is robust, and the
error-bars are realistic. Because the second two templates were of
better quality, we adopted the SME values found from these spectra with
simple averaging, yielding $\teffstar=\hatcurSMEteff$\,K, $\feh =
\hatcurSMEzfeh$ and $\vsini = \hatcurSMEvsin$\,\kms. We adopted
these as the final atmospheric parameters for the star.
We then also repeated the isochrone search for stellar parameters,
obtaining \mstar=\hatcurISOmlong\,\msun, \rstar=\hatcurISOrlong\,\rsun\
and \lstar=\hatcurISOlum\,\lsun.  These are summarized in
\reftab{stellar}, along with other stellar properties. Model isochrones
from \citet{yi:2001} for metallicity \feh=\hatcurSMEzfehshort\ are
plotted in \reffig{iso}, with the final choice of effective temperature
$\teffstar$ and \arstar\ marked, and encircled by the 1-$\sigma$ and
2-$\sigma$ confidence ellipsoids. The second SME iteration at fixed
stellar surface gravity (as determined from \arstar) changed the
metallicity and stellar temperature in such a way that the new
(\teffstar, \arstar) values now fall in a more complex region of the
isochrones, as compared to the initial SME values, allowing for
multiple solutions (see \reffig{iso}, where the original SME values are
marked with a triangle).\footnote{The reason for the intersecting
isochrones is the ``kink'' on the evolutionary tracks for
$\mstar\gtrsim1.0\,\msun$ stars evolving off the main sequence. }
%
%
The distribution of stellar age becomes bimodal with the dominant peak
in the histogram (not shown) being at 5.0\,Gyr, and a smaller peak (by
a factor of 5) at 7.3\,Gyr. This corresponds to a slightly bimodal mass
distribution with the dominant peak at $\sim 1.23$\,\msun, and much
smaller peak around $\sim 1.13\,\msun$. The asymmetric error-bars given
in \reftab{stellar} for the mass and age account for the
double-peaked distribution.

The stellar evolution modeling also yields the absolute magnitudes and
colors in various photometric passbands. We used the apparent
magnitudes from the 2MASS catalogue \citep{skrutskie:2006} to determine
the distance of the system, after conversion to the ESO system of the
models. 
The reported magnitudes for this star are
$J_{\rm 2MASS}=\hatcurCCtwomassJmag$, 
$H_{\rm 2MASS}=\hatcurCCtwomassHmag$ and 
$K_{\rm 2MASS}=\hatcurCCtwomassKmag$,
which we transformed to
$J_{\rm ESO}=\hatcurCCesoJmag$, 
$H_{\rm ESO}=\hatcurCCesoHmag$ and 
$K_{\rm ESO}=\hatcurCCesoKmag$, 
following \citet[see][]{carpenter:2001}. These yield a color of
$(J-K)=0.378\pm0.028$, fully consistent with the expected,
isochrone-based $(J-K)_{\rm iso}=0.374\pm0.015$. We thus relied on the
2MASS $K$ apparent magnitude and the $M_{\rm K}=\hatcurISOMK$ absolute
magnitude derived from the above-mentioned modeling to determine the
distance: \hatcurXdist\,pc, assuming no reddening. The \band{K} was
chosen because it is the longest wavelength band-pass with the smallest
expected discrepancies due to molecular lines in the spectrum of the
star.

\newcommand{\hatcurisoshort}{YY}
\newcommand{\hatcurlumind}{\arstar}
\begin{deluxetable}{lcl}
\tablewidth{0pc}
\tablecaption{
	Stellar parameters for \hatcur{}
	\label{tab:stellar}
}
\tablehead{
	\colhead{Parameter}	&
	\colhead{Value} &
	\colhead{Source}
}
\startdata
$\teffstar$ (K)\dotfill         &  \hatcurSMEteff   & SME\tablenotemark{a}\\
$\feh$ (dex)\dotfill            &  \hatcurSMEzfeh   & SME                 \\
$\vsini$ (\kms)\dotfill         &  \hatcurSMEvsin   & SME                 \\
$\vmac$ (\kms)\dotfill          &  \hatcurSMEvmac   & SME                 \\
$\vmic$ (\kms)\dotfill          &  \hatcurSMEvmic   & SME                 \\
$\gamma_{\rm RV}$ (\kms)\dotfill         &  \hatcurDSgamma       & DS      \\
$\gamma_{1(i)}$\dotfill			&  \hatcurLBii		& SME+Claret          \\
$\gamma_{2(i)}$\dotfill			&  \hatcurLBiii		& SME+Claret          \\
$\mstar$ ($\msun$)\dotfill      &  \hatcurISOm       & \hatcurisoshort+\hatcurlumind+SME\tablenotemark{c}\\
$\rstar$ ($\rsun$)\dotfill      &  \hatcurISOr       & \hatcurisoshort+\hatcurlumind+SME         \\
$\loggstar$ (cgs)\dotfill       &  \hatcurISOlogg    & \hatcurisoshort+\hatcurlumind+SME         \\
$\lstar$ ($\lsun$)\dotfill      &  \hatcurISOlum     & \hatcurisoshort+\hatcurlumind+SME         \\
$V$ (mag)\dotfill				&  \hatcurCCtassmv   & TASS\tablenotemark{d}                     \\
$B-V$ (mag)\dotfill				&  $0.73\pm0.03$     & \hatcurisoshort+\hatcurlumind+SME         \\
$M_V$ (mag)\dotfill             &  \hatcurISOmv      & \hatcurisoshort+\hatcurlumind+SME         \\
$K$ (mag,ESO)                   &  \hatcurCCcitKmag  & 2MASS\tablenotemark{e}                    \\
$M_K$ (mag,ESO)\dotfill         &  \hatcurISOMK      & \hatcurisoshort+\hatcurlumind+SME         \\
Age (Gyr)\dotfill               &  \hatcurISOage     & \hatcurisoshort+\hatcurlumind+SME         \\
Distance (pc)\dotfill           &  \hatcurXdist      & \hatcurisoshort+\hatcurlumind+SME         \\
\enddata
\tablenotetext{a}{
	SME = ``Spectroscopy Made Easy'' package for analysis of
	high-resolution spectra \cite{valenti:1996}. These parameters
	depend primarily on SME, with a small dependence on the iterative
	analysis incorporating the isochrone search and global modeling of
	the data, as described in the text.
}
\tablenotetext{b}{
	SME+Claret = Based on SME analysis and tables from
	\citet{claret:2004}.
}
\tablenotetext{c}{
	\hatcurisoshort+\hatcurlumind+SME = \citet{yi:2001} isochrones,
	\arstar\ luminosity indicator, and SME results.
}
\tablenotetext{d}{
	Based on the TASS catalogue \citep{droege:2006}.
}
\tablenotetext{e}{Based on the transformations from \citet{carpenter:2001}.}
\end{deluxetable}

\subsection{Excluding blend scenarios}
\label{sec:blend}

\label{sec:bisec}

Following \cite{torres:2007}, we explored the possibility that the
measured radial velocities are not due to the (multiple) planet-induced
orbital motion of the star, but are instead caused by distortions in
the spectral line profiles. This could be due to contamination from a nearby
unresolved eclipsing binary, in this case presumably with a second
companion producing the RV signal corresponding to \hatcurc. A
bisector analysis based on the Keck spectra was done as described in \S
3 of \cite{hartman:2009}.

We detect no significant variation in the bisector spans (see
\reffig{rv}, fifth panel). The correlation between the radial velocity
and the bisector variations is also insignificant. Therefore, we
conclude that the velocity variations of the host star are real, and
can be interpreted as being due to a close-in planet, with the added
complication from an outer object that we account for in the modeling
that follows.  Because of the negligible bisector variations that show
no correlation with the radial velocities, we found no need to perform
detailed blend modeling of hierarchical triple (quadruple) scenarios,
such as that done for the case of HAT-P-12b \citep{hartman:2009}.

\subsection{Global modeling of the data}
\label{sec:globmod}

This section presents a simultaneous fitting of the HATNet photometry,
the follow-up light curves, and the RV observations, which we refer to
as ``global'' modeling. It incorporates not only a physical model of
the system, but also a description of systematic (instrumental)
variations.

Our model for the follow-up \lcs\ used analytic formulae based on
\citet{mandel:2002} for the eclipse of a star by a planet, where the
stellar flux is described by quadratic limb-darkening. The limb
darkening coefficients were taken from the tables by
\citet{claret:2004} for the \band{i}, corresponding to the stellar
properties determined from the SME analysis (\refsec{stelparam}). The
transit shape was parametrized by the normalized planetary radius
$p\equiv \rpl/\rstar$, the square of the impact parameter $b^2$, and
the reciprocal of the half duration of the transit $\zrstar$. We chose
these parameters because of their simple geometric meanings and the
fact that they show negligible correlations \citep[see
e.g.][]{carter:2008}.

Our model for the HATNet data was the simplified ``P1P3'' version of
the \citet{mandel:2002} analytic functions (an expansion by Legendre
polynomials), for the reasons described in \citet{bakos:2009}.
The depth of the HATNet transits was adjusted independently in the fit (the
depth was $B_{inst}$ ``blending factor'' times that of the follow-up
data) to allow for possible contamination by nearby stars in the
under-sampled images of HATNet.

As indicated earlier, initial modeling of the RV observations showed
deviations from a Keplerian fit highly suggestive of a second body in
the system with a much longer period than the transiting planet. Thus,
in our global modeling, the RV curve was parametrized by the
combination of an eccentric Keplerian orbit for the inner planet with
semi-amplitude $K$, and Lagrangian orbital elements
$(k,h)=e\times(\cos\omega,\sin\omega)$, plus an eccentric Keplerian
orbit for the outer object with $K_2$, $k_2$ and $h_2$, and a systemic
RV zero-point $\gamma$. Throughout this paper the subscripts ``1'' and
``2'' will refer to \hatcurb\ and \hatcurc, respectively. If the
subscript is omitted, we refer to \hatcurb.

In the past, for single transiting planet scenarios we have assumed
strict periodicity in the individual transit times \citep[][ and
references therein]{hartman:2009}, even when a drift in the RV
measurements indicated an outer companion
\citep[HAT-P-11b,][]{bakos:2009}. Since the expected transit timing
variations (TTV) for these planets were negligible compared to the
error-bars of the transit center measurements, the strict periodicity
was a reasonable hypothesis. Those data were characterized by two
transit centers ($T_A$ and $T_B$), and all intermediate transits were
interpolated using these two epochs and their corresponding $N_{tr}$
transit numbers. The model for the RV data component also implicitly
contained its ephemeris information through $T_A$ and $T_B$, and thus
was coupled with the photometry data in the time domain.

For \hatcurb, however, the disturbing force by the outer planet
\hatcurc\ is expected to be not insignificant, because the RV
semi-amplitude of the host star ($\sim 0.5\,\kms$) indicates that
\hatcurc\ is massive, and the relatively short period and eccentric
orbit (see later) indicate that it moves in relatively close to
\hatcurb. Thus, the assumption of strict periodicity for \hatcurb\ is
not precisely correct. While we performed many variations of the global
modeling,
in our finally adopted physical model we assume strict periodicity only
for the HATNet data, where the timing error on individual transits can
be $\sim 1000$\,seconds. In this final model we allow for departure
from such a periodicity for the individual transit times for the seven
follow-up photometry observations. In practice, we assigned the transit
number $N_{tr} = 0$ to the first complete high quality follow-up \lc\ gathered
on 2008 November 8/9 (MST). The HATNet data-set was characterized by
$T_{c,-370}$ and $T_{c,-309}$, covering all transit events observed by
HATNet (here the $c$ subscript denotes ``center'' for the transits of
\hatcurb).  The transit follow-up observations were characterized by their
respective times of transit center: $T_{c,-68}$, $T_{c,-1}$, $T_{c,0}$,
$T_{c,1}$, $T_{c,24}$, $T_{c,35}$, $T_{c,62}$. Initial estimates of the
$T_{c,i}$ values yielded an initial epoch $T_c$ and period $P$ by
linear fitting weighted by the respective error-bars of the transit
centers.  The model for the RV data component of the inner planet
contained the ephemeris information through the $T_c$ and $P$
variables, i.e.~it was coupled with the transit data. The global
modeling was done in an iterative way. After an initial fit to the
transit centers (and other parameters; see later), $T_c$ and $P$ were
refined, and the fit was repeated.

The time dependence of the RV of the outer planet was described via its
hypothetical transit time $T_{2c}$ (as if its orbit were edge on), and
its period $P_2$.  The time of periastron passage $T_{2peri}$ can be
equivalently used in place of time of conjunction $T_{2c}$. 


Altogether, the 21 parameters describing the physical model were
$T_{c,-370}$, $T_{c,-309}$ (HATNet), $T_{c,-68}$, $T_{c,-1}$,
$T_{c,0}$, $T_{c,1}$, $T_{c,24}$, $T_{c,35}$, $T_{c,62}$ (\flwof),
$\rpl/\rstar$, $b^2$, $\zrstar$, $K$, $\gamma$, $k = e\cos\omega$, $h =
e\sin\omega$, $K_2$, $k_2$, $h_2$, $P_2$ and $T_{2c}$. Two additional
auxiliary parameters were the instrumental blend factor $B_{inst}$ of
HATNet, and the HATNet out-of-transit magnitude, $M_{\rm 0,HATNet}$.

\begin{figure}[!ht]
\plotone{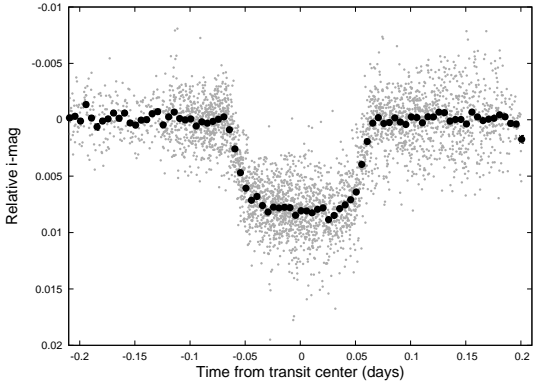}
\caption{
	Part of the global modeling described in \refsec{analysis} are
	corrections for systematic variations of the \lcs\ via simultaneous
	fitting with the physical model of the transit. The figure shows
	the resulting EPD- and TFA-corrected \lcs\ for all 7 follow-up
	events (small gray points), and the merged and binned \lc\
	(bin-size 0.005\,days).
\label{fig:phfuall}}
\end{figure}

We extended our physical model with an instrumental model that
describes the systematic (non-physical) variations (such as trends) of
the follow-up data. This was done in a similar fashion to
the analysis presented in \citet{bakos:2009}. The HATNet photometry had
already been EPD- and TFA-corrected before the global modeling, so we
only modeled systematic effects in the follow-up \lcs. We chose the
``ELTG'' method, i.e.~EPD was performed in ``local'' mode with EPD
coefficients defined for each night, whereas TFA was performed in
``global mode'', with the same coefficients describing the optimal
weights for the selected template stars in the field.  The five EPD
parameters include the hour angle (characterizing a monotonic trend
that changes linearly over a night), the square of the hour angle, and
three stellar profile parameters (equivalent to FWHM, elongation, and
position angle). The exact functional form of the above parameters
contained 6 coefficients, including the auxiliary out-of-transit
magnitude of the individual events. The EPD parameters were independent
for all 7 nights, implying 42 additional coefficients in the global
fit.  For the global TFA analysis we chose 19 template stars that had
good quality measurements for all nights and on all frames, implying 19
additional parameters in the fit.  Thus, the total number of fitted
parameters was 21 (physical model) + 2 (auxiliary) + 42 (local EPD) +
19 (TFA) = 84.


The joint fit was performed as described in \citet{bakos:2009}, by
minimizing \chisq\ in the parameter space using a hybrid algorithm,
combining the downhill simplex method \citep[AMOEBA, see][]{press:1992}
with the classical linear least-squares algorithm.  
We used the partial derivatives of the model functions as
derived by \citet{pal:2008b}. Uncertainties in the parameters were
derived using the Markov Chain Monte-Carlo method \citep[MCMC,
see][]{ford:2006}.
%
%
Since the eccentricity of the inner system appeared as marginally
significant ($k=\hatcurRVk$, $h=\hatcurRVh$, implying
$e=\hatcurRVeccen$), and also because the physical model dictates that
in the presence of a massive outer companion, the inner planet could
maintain a non-zero eccentricity, we did not fix $k$ and $h$ to zero.
%
%
The best fit results for the relevant physical parameters are
summarized in \reftab{planetparam} and \reftab{planetparamc}. The
individual transit centers for the \flwof\ data are given in
\refsec{ttv}. Other parameters fitted but not listed in the table were:
$T_{\mathrm{c},-370}=\hatcurLCTA$~(BJD), 
$T_{\mathrm{c},-309}=\hatcurLCTB$~(BJD), 
$B_{inst}=\hatcurLCiblend$,
and $M_{\rm 0,HATNet}=\hatcurLChatnetm$ (\band{I}).
The fit to the HATNet photometry data was shown earlier in
\reffig{hatnet}, the orbital fit to the RV data is shown in
\reffig{rv}, and the fit to the \flwof\ data is displayed in
\reffig{lc}. The low rms of 3.4\,\ms\ around the orbital fit is due in
part to the use of an iodine free Keck/HIRES template that was acquired
with higher spectral resolution and higher S/N than usual. Note that
the low rms implies the absence of any additional interior planets
other than \hatcurb, consistent with expectations given that the
massive and eccentric outer planet \hatcurc\ dynamically forbids such
planets \citep[see e.g.][]{wittenmyer:2007}. 
The stellar jitter required to reconcile the reduced \chisq\ 
with $1.0$ for the RV data is \hatcurRVjitter\,\ms. The low jitter is
also consistent with \hatcur\ being a chromospherically quiet star
(based on \logrhk\ and $S$).

The planetary parameters and their uncertainties can be derived by the
direct combination of the \emph{a posteriori} distributions of the \lc,
radial velocity and stellar parameters.  We find that the mass of the
inner planet is $\mpl=\hatcurPPmlong\,\mjup$, the radius is
$\rpl=\hatcurPPrlong\,\rjup$, and its density is
$\rho_p=\hatcurPPrho$\,\gcmc.  The final planetary parameters are
summarized at the bottom of Table~\ref{tab:planetparam}. The
simultaneous EPD- and TFA-corrected \lc\ of all photometry follow-up
events is shown in \reffig{phfuall}.

\begin{figure}[!ht]
\plotone{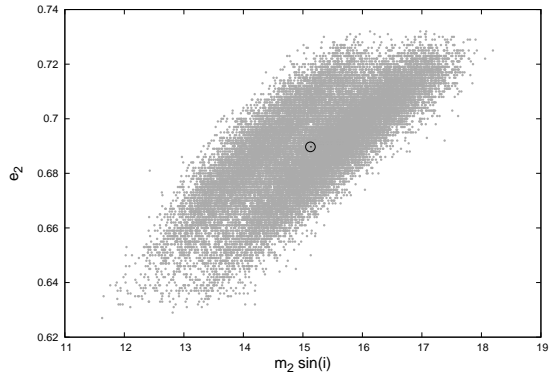}
\caption{
%
	The orbital parameters, and as a consequence, the mass ($m_2\sin
	i$) and eccentricity $e_2$ of the outer planet are strongly
	correlated (see \refsec{analysis}). Gray dots represent the results
	of the MCMC runs (20,000 trials). The final solution is marked with
	an open circle. The bimodal distribution in $m_2\sin i$ is due to
	the similar distribution of the stellar mass (see
	\refsec{stelparam}).
\label{fig:deg}}
\end{figure}

The outer planet, \hatcurc, appears to be very massive, with $m_2\sin
i_2 = \hatcurcPPmlong\,\mjup$, and orbits the star in a highly
eccentric orbit with $e_2=\hatcurcRVeccen$. The orientation of the
orbit ($\omega_2 = \hatcurcRVomega\arcdeg$) is such that our line of
sight is almost along the minor axis, coincidentally as in the HAT-P-2b
system \citep{bakos:2007}. We note that the periastron passage of
\hatcurc\ has not been well monitored, and thus the RV fit of the orbit
suffers from a strong correlation between the quantities $K_2$, $e_2$
and $\gamma$, leading to correlated $m_2\sin i$ and $e_2$ (\reffig{deg}).

\begin{deluxetable}{lr}
\tablewidth{0pc}
\tablecaption{Orbital and planetary parameters for \hatcur\lowercase{b}
	\label{tab:planetparam}}
\tablehead{
	\colhead{~~~~~~~~~~~~~~~Parameter~~~~~~~~~~~~~~~} &
	\colhead{Value}
}
\startdata
\sidehead{\Lc{} parameters, \hatcurb}
~~~$P$ (days)             \dotfill    & $\hatcurLCP$              \\
~~~$T_c$ (${\rm BJD}$)    \dotfill    & $\hatcurLCT$              \\
~~~$T_{14}$ (days)
      \tablenotemark{a}   \dotfill    & $\hatcurLCdur$            \\
~~~$T_{12} = T_{34}$ (days)
    \tablenotemark{a}     \dotfill    & $\hatcurLCingdur$         \\
~~~$\arstar$              \dotfill    & $\hatcurPPar$             \\
~~~$\zrstar$              \dotfill    & $\hatcurLCzeta$           \\
~~~$\rpl/\rstar$          \dotfill    & $\hatcurLCrprstar$        \\
~~~$b^2$                  \dotfill    & $\hatcurLCbsq$            \\
~~~$b \equiv a \cos i/\rstar$
                          \dotfill    & $\hatcurLCimp$            \\
~~~$i$ (deg)              \dotfill    & $\hatcurPPi$ \phn         \\
~~~$T_{peri}$ (days)      \dotfill    & $\hatcurPPperi$           \\

\sidehead{RV parameters, as induced by \hatcurb}
~~~$K$ (\ms)              \dotfill    & $\hatcurRVK$              \\
~~~$k$                    \dotfill    & $\hatcurRVk$              \\
~~~$h$                    \dotfill    & $\hatcurRVh$              \\
~~~$e$                    \dotfill    & $\hatcurRVeccen$          \\
%
~~~$\omega$               \dotfill    & $\hatcurRVomega^\circ$   \\

\sidehead{Other RV parameters}
~~~$\gamma_{rel}$ (\ms)   \dotfill    & $\hatcurRVgammarel$       \\
~~~Jitter (\ms)           \dotfill    & \hatcurRVjitter           \\

\sidehead{Secondary eclipse parameters for \hatcurb}
~~~$T_s$ (BJD)            \dotfill    & $\hatcurXsecondary$       \\
~~~$T_{s,14}$             \dotfill    & $\hatcurXsecdur$          \\
~~~$T_{s,12}$             \dotfill    & $\hatcurXsecingdur$       \\

\sidehead{Planetary parameters for \hatcurb}
~~~$\mpl$ ($\mjup$)       \dotfill    & $\hatcurPPmlong$          \\
~~~$\rpl$ ($\rjup$)       \dotfill    & $\hatcurPPrlong$          \\
~~~$C(\mpl,\rpl)$
    \tablenotemark{b}     \dotfill    & $\hatcurPPmrcorr$         \\
~~~$\rhopl$ (\gcmc)       \dotfill    & $\hatcurPPrho$            \\
~~~$a$ (AU)               \dotfill    & $\hatcurPParel$           \\
~~~$\log g_p$ (cgs)       \dotfill    & $\hatcurPPlogg$           \\
~~~$T_{\rm eq}$ (K)       \dotfill    & $\hatcurPPteff$           \\
~~~$\Theta$ \tablenotemark{c}              
                          \dotfill    & $\hatcurPPtheta$          \\
%
~~~$F_{per}$ (\ergscmsq) \tablenotemark{d}
                          \dotfill    & $\hatcurPPfluxperi$      \\
~~~$F_{ap}$  (\ergscmsq) \tablenotemark{e} 
                          \dotfill    & $\hatcurPPfluxap$        \\
~~~$\langle F \rangle$ (\ergscmsq)
                          \dotfill    & $\hatcurPPfluxavg$        \\
\enddata
\tablenotetext{a}{
	\ensuremath{T_{14}}: total transit duration, time
		between first to last contact;
	\ensuremath{T_{12}=T_{34}}: ingress/egress time, time between first
		and second, or third and fourth contact.
}
\tablenotetext{b}{
	Correlation coefficient between the planetary mass \mpl\ and radius
	\rpl.
}
\tablenotetext{c}{
	The Safronov number is given by $\Theta =
	\frac{1}{2}(V_{\rm esc}/V_{\rm orb})^2 = (a/\rpl)(\mpl / \mstar )$
	\citep[see][]{hansen:2007}.
}
\tablenotetext{d}{
	Incoming flux per unit surface area.
%
}
\end{deluxetable}

\begin{deluxetable}{lr}
\tablewidth{0pc}
\tablecaption{Orbital and planetary parameters for \hatcur\lowercase{c}
\label{tab:planetparamc}}
\tablehead{
	\colhead{~~~~~~~~~~~~~~~Parameter~~~~~~~~~~~~~~~} &
	\colhead{Value}
}
\startdata
\sidehead{RV parameters, as induced by \hatcurc}
~~~$P_2$ (days)             \dotfill    & $\hatcurcLCP$              \\
~~~$T_{2c}$\tbn{a} (BJD)    \dotfill    & $\hatcurcLCT$              \\
~~~$K_2$ (\ms)              \dotfill    & $\hatcurcRVK$              \\
~~~$k_2$                    \dotfill    & $\hatcurcRVk$              \\
~~~$h_2$                    \dotfill    & $\hatcurcRVh$              \\
~~~$e_2$                    \dotfill    & $\hatcurcRVeccen$          \\
~~~$\omega_2$               \dotfill    & $\hatcurcRVomega^\circ$    \\
~~~$T_{2,peri}$ (days)      \dotfill    & $\hatcurcPPperi$           \\

\sidehead{Fictitious \lc{} parameters, \hatcurc\tablenotemark{b}}
~~~$T_{2,14}$\tbn{c} (days)   \dotfill    & $\hatcurcLCdur$          \\
~~~$T_{2,12} = T_{34}$ (days) \dotfill    & $\hatcurcLCingdur$       \\

\sidehead{Fictitious secondary eclipse parameters for \hatcurc\tbn{a}}
~~~$T_{2s}$ (BJD)          \dotfill    & $\hatcurcXsecondary$        \\
~~~$T_{2s,14}$ (days)      \dotfill    & $\hatcurcXsecdur$           \\
~~~$T_{2s,12}$ (days)      \dotfill    & $\hatcurcXsecingdur$        \\

\sidehead{Planetary parameters for \hatcurc}
~~~$m_2\sin i_2$ ($\mjup$) \dotfill    & $\hatcurcPPmlong$           \\
~~~$a_2$ (AU)              \dotfill    & $\hatcurcPParel$            \\
~~~$T_{2,\rm eq}$ (K)      \dotfill    & $\hatcurcPPteff$            \\
%
~~~$F_{2,per}$ (\ergscmsq) \tbn{d}
                          \dotfill    & $\hatcurcPPfluxperi$         \\
~~~$F_{2,ap}$  (\ergscmsq) \tbn{d} 
                          \dotfill    & $\hatcurcPPfluxap$           \\
~~~$\langle F_2 \rangle$ (\ergscmsq) \tbn{d}
                          \dotfill    & $\hatcurcPPfluxavg$          \\

\enddata
\tablenotetext{a}{
	$T_{2c}$ would be the center of transit of \hatcurc, if its
	(unknown) inclination was $90\arcdeg$.
}
\tablenotetext{b}{
	Transits of \hatcurc\ have not been observed. The values are for
	guidance only, and assume zero impact parameter.  
}
\tablenotetext{c}{
	\ensuremath{T_{14}}: total transit duration, time
	between first to last contact, assuming zero impact parameter. 
	\ensuremath{T_{12}=T_{34}}: ingress/egress time, time between first
	and second, or third and fourth contact.  Note that these values
	are fictitious, and transits of \hatcurc\ have not been observed.
}
\tablenotetext{d}{
	Incoming flux per unit surface area in periastron, apastron, and
	averaged over the orbit. 	
}
\end{deluxetable}
\vspace{10mm}


\section{Discussion}
\label{sec:discussion}

We present the discovery of HAT-P-13, the first detected multi-planet
system with a transiting planet.  The inner transiting planet,
\hatcurb, is an inflated ``hot Jupiter'' in a nearly circular orbit.
The outer planet, \hatcurc, is both extremely massive and highly
eccentric, and orbits in a $P>1$ yr orbit . With an iron abundance of
$\feh = \hatcurSMEzfeh$, the host star is also remarkable.  As we
describe below, this extraordinary system is a rich dynamical
laboratory, the first to have an accurate clock (\hatcurb) and a known
perturbing force (\hatcurc). The inner planet will help refine the
orbital configuration (and thus true mass) of the outer planet through
transit timing variations (TTVs). Conversely, the outer planet, through
its known perturbation, may constrain structural parameters and the
tidal dissipation rate of the inner planet \citep{batygin:2009}, in
addition to all the information that can be gleaned from the transits
of \hatcurb.

\subsection{The planet \hatcurb}
\label{sec:hatcurb}

The only other known planet with a host-star as metal rich as \hatcur\
(\feh=\hatcurSMEzfeh) is XO-2b \citep[$\mathrm \lbrack Fe/H\rbrack =
+0.45$,][]{burke:2007}.  XO-2b, however, has much smaller mass
(0.56\,\mjup) and smaller radius (0.98\,\rjup) than \hatcurb\
(\hatcurPPmshort\,\mjup, \hatcurPPrshort\,\rjup). Other planets similar
to \hatcurb\ include HAT-P-9b \citep[$\mpl=0.78\,\mjup$,
$\rpl=1.4\,\rjup$, ][]{shporer:2008}, and XO-1b
\citep[$\mpl=0.92\,\mjup$, $\rpl=1.21\,\rjup$, ][]{pmcc:2006}.

When compared to theoretical models, \hatcurb\ is a clearly inflated
planet. The \citet{baraffe:2008} models with solar insolation (at
0.045\,AU) are consistent with the observed mass and radius of
\hatcurb\ either with overall metal content Z=0.02 and a very young
age of 0.05--0.1\,Gyr, or with metal content Z=0.1 and age
0.01--0.05\,Gyr.\footnote{The equivalent semi-major axis, at which
\hatcurb\ would receive the same amount of insolation when orbiting our
Sun, is $a_{equiv} = \hatcurPPaequiv$\,AU.} Given the fact, however,
that the host star is metal rich, and fairly old (\hatcurISOage\,Gyr),
it is unlikely that \hatcurb\ is newly formed (50\,Myr) and very
metal poor. Comparison with \citet{fortney:2007} leads to
similar conclusions. Using these models, \hatcurb\ is broadly
consistent only with a 300\,Myr planet at 0.02\,AU solar distance and
core-mass up to 25\,\mearth, or a 1\,Gyr core-less (pure H/He) planet. 
If \hatcurb\ has a significant rocky core, consistent with
expectation given the high metallicity of \hatcur, it must somehow be
inflated beyond models calculated for such a planet with age and
insolation suggested by the data. Numerous explanations have been
brought up in the past to explain the inflated radii of certain
extrasolar planets \citep[][and references
therein]{miller:2009,fabrycky:2007}. One among these has been the tidal
heating due to non-zero eccentricity \citep{bodenheimer:2001}. No
perturbing companion has been found for the inflated planets
\hd{209458b} and HAT-P-1b \citep[for an overview,
see][]{mardling:2007}. The \hatcur\ system may be the first, where the
inflated radius can be explained by the non-zero eccentricity of
\hatcurb, excited by the outer \hatcurc\ planet orbiting on a highly
eccentric orbit. We note that while the eccentricity of \hatcurb\ is
non-zero only at the 2-$\sigma$ level (because $k$ is non-zero at
2-$\sigma$), its pericenter is aligned (to within
$4\arcdeg\pm40\arcdeg$) with that of the outer planet \hatcurc.
%
Additional RV measurements and/or space-based timing of a secondary
eclipse are necessary for a more accurate determination of the orbit.
Nevertheless, if the apsidal lines of \hatcurb\ and \hatcurc\ are indeed
aligned, and, if the configuration is co-planar, then the system is in
a tidal fix-point configuration, as recently noted by
\citet{batygin:2009}. This configuration imposes constraints on the
structure of the inner planet \hatcurb. In particular,
\citet{batygin:2009} give limits on the tidal Love number $k_{2b}$,
core-mass and tidal energy dissipation rate $Q_b$ of \hatcurb. 

\subsection{Transit timing variations}
\label{sec:ttv}

It has long been known that multiple planets in the same planetary
system perturb each other, and this may lead to detectable variations
in the transit time of the transiting planet(s). Transit timing
variations have been analytically described by \citet{holman:2005} and
\citet{agol:2005}, and have been suggested as one of the most effective
ways of detecting small mass perturbers of transiting planets. This has
motivated extensive follow-up of known TEPs e.g.~by the Transit Light
Curve (TLC) project \citep{holman:2006,winn:2007}, looking for
companions of TEPs. However, for \hatcurb\ there is observational
(spectroscopic) evidence for an outer component (\hatcurc), and thus
transit timing variations of \hatcurb\ must be present. Transit timing
variations can be used to constrain the mass and orbital elements of
the perturbing planet, in our case those of \hatcurc.

\begin{figure}[!ht]
\plotone{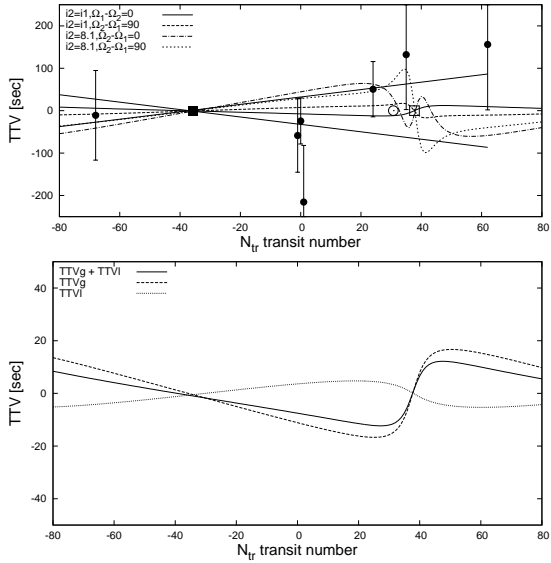}
\caption{
	(Top) Transit times of the individual transit events of \hatcurb.
	The filled circles correspond to the global analysis described in
	\refsec{ttv}. The large filled square, the open circle and the
	open square correspond to the apastron, conjunction and periastron
	of the perturber \hatcurc. Overlaid are analytic models for the
	transit timing variation due to gravitational effects (TTVg) and
	light-travel time effects (TTVl) for four scenarios: i) the
	inclination of \hatcurc\ is $i_2 = 83.1\arcdeg = i_1$, ii)
	$i_2=8.1\arcdeg$ (at fixed $m_2\sin i_2$, corresponding to a large
	mass for \hatcurc), iii) $i_2 = 83.1\arcdeg$ and the mutual
	inclination of the orbital normals in the plane of the sky is $D =
	\Omega_1 - \Omega_2 = 90\arcdeg$ (see Fig.~1 of
	\citet{borkovits:2003}), or iv) $i_2 = 8.1\arcdeg$ and $D =
	90\arcdeg$.
	(Bottom) The selected case ``i'' from above with zoomed-in vertical
	scale, showing the TTVg (dashed line) and TTVl (dotted line)
	effects separately, and their net effect (solid line).
\label{fig:ttv}}
\end{figure}

The global modeling described in \refsec{globmod} treats the transit
centers of the \flwof\ telescope follow-up as independent variables,
i.e., it automatically provides the basis for a TTV analysis. 
Throughout this work, by TTV we refer to the time difference between
the observed transit center ($T_{c,i}$) and the calculated value based
on a fixed epoch and period as given in \reftab{planetparam},
i.e.~in the $O-C$ sense. The resulting TTVs are shown in \reffig{ttv}. 
We believe the error-bars to be realistic as they are the result of a
full MCMC analysis, where all parameters are varied (including the EPD
and TFA parameters). As further support for this, the transits around
$N_{tr} = 0$ ($T_{c,-1}$,$T_{c,0}$,$T_{c,1}$) show a standard deviation
that is comparable to the error-bars (and we can safely assume zero TTV
within a $\pm\hatcurLCPshort$~day time range). It is also apparent from
the plot that the smallest error-bars correspond to the full transit
observations ($N_{tr}=$ 0 and 24). TTVs of the order of 100 seconds are
seen from the best fit period. Given the large error-bars on our
transit centers (of the order of 100 seconds), in part due to possible
remaining systematics in the partial transit \lc\ events, we consider
these departures suggestive only.

Nevertheless, it is tempting to compare our results with analytic
formulae presented by \citet{agol:2005} and \citet[][]{borkovits:2003}. 
The \hatcur(b,c) system corresponds to case ``ii'' of
\citet{agol:2005}, i.e.~with an exterior planet on an eccentric orbit
having a much larger semi-major axis than the inner planet.  Formulae
for the general (non co-planar) case are given by
\citet[][Eq.~46]{borkovits:2003}. The TTV effect will depend on the
following known parameters for the \hatcur\ system: $\mstar$, $\mpl$,
$P$, $\ipl$, $P_2$, $e_2$, $\omega_2$, $m_2\sin i_2$ and $T_{2,peri}$
(these are listed in \reftab{planetparam} and \reftab{planetparamc}).
The TTV will also depend on the following unknown parameters: the true
inclination of \hatcurc, $i_2$, and the relative angle of the orbital
normals projected onto the plane of the sky, $D = \Omega_1 - \Omega_2$
\citep[see Fig.~1 of ][]{borkovits:2003}. In addition to the
gravitational perturbation of \hatcurc\ on \hatcurb, the barycenter of
the inner subsystem (composed of the host star \hatcur, and the inner
planet \hatcurb) orbits about the three-body barycenter due to the
massive \hatcurc\ companion. This leads to light-travel time effects in
the transit times of \hatcurb\ (TTVl) that are of the same order of
magnitude as the TTV effect due to the perturbation (TTVg).

We have evaluated the analytic formulae including both the TTVg and
TTVl effects for cases with $i_2 = \hatcurPPi\arcdeg$ (corresponding to
co-planar inner and outer orbits, and to $M_2 =
\hatcurcPPmshort\,\mjup$), and $i_2 = 8.1\arcdeg$ (an ad hoc value
yielding an almost face-on orbit with $M_2 = 105\,\mjup$), 
and $D = 0\arcdeg$ or $D = 90\arcdeg$.
These four analytic models are illustrated in \reffig{ttv}. The bottom
panel of \reffig{ttv} shows the TTVl and TTVg effects separately for
the $i_2 = \hatcurPPi\arcdeg$ and $D = 0\arcdeg$ case.  The $i_2 =
\hatcurPPi\arcdeg$ ($M_2 = \hatcurcPPmshort\,\mjup$) cases give TTV
variations of the order of 15 seconds. The functional form of the $i_2
= \hatcurPPi\arcdeg$ and $D = 0\arcdeg$ case appears to follow the
observational values, albeit with much smaller amplitude. Curiously,
for this case the TTVl effect cancels the TTVg effect to some extent
(bottom panel of \reffig{ttv}). Increasing the mass of the outer
companion by decreasing $i_2$ at constant $m_2\sin i_2$ does increase
the TTV amplitude up to 100 seconds at $i_2 = 8.1\arcdeg$, but the
functional form changes and no longer resembles the trend seen in the
observational data.

In conclusion, while it is premature to fit the current data-set with
analytic models because of the small number of data-points (7) and the
large error-bars ($\sim100$~sec), these data are not inconsistent with
the presence of TTVs, and will prove useful in later analyses. Future
measurements of full transit events with high accuracy in principle can
determine both $i_2$ and $D$, i.e., \hatcurc\ may become an RV-based
detection with known orbital orientation and a true mass (even if it
does not transit).

\begin{deluxetable}{lrrr}
\tablewidth{0pc}
\tablecaption{
	Transit timing variations of \hatcur{}.
	\label{tab:ttv}
}
\tablehead{
	\colhead{$N_{tr}$} & 
	\colhead{$T_c$} & 
	\colhead{\ensuremath{\sigma_{\rm T_c}}} & 
	\colhead{O-C}\\
	\colhead{} & 
	\colhead{(BJD)} & 
	\colhead{(sec)} &
	\colhead{(sec)}
}
\startdata
$ -68 $ & $ 2454581.62406 $ & $ 105.7 $ & $ -10.9 $\\
$ -1 $ & $ 2454777.01287 $ & $ 86.7 $ & $ -58.5 $\\
$ 0 $ & $ 2454779.92953 $ & $ 54.3 $ & $ -24.2 $\\
$ 1 $ & $ 2454782.84357 $ & $ 133.6 $ & $ -215.6 $\\
$ 24 $ & $ 2454849.92062 $ & $ 65.1 $ & $ 50.7 $\\
$ 35 $ & $ 2454882.00041 $ & $ 129.5 $ & $ 132.2 $\\
$ 62 $ & $ 2454960.73968 $ & $ 154.1 $ & $ 156.1 $\\
\enddata
\end{deluxetable}

\subsection{The planet \hatcurc}
\label{sec:hatcurc}

The probability of \hatcurc\ transiting the host star, as seen from the
Earth, depends on \rstar, $R_{2p}$, $e_2$, $\omega_2$, and $a_2$
\citep{kane:2009}. We evaluated the transit probability in a Monte
Carlo fashion, as part of the global modeling, resulting in $P_{2,tr} =
0.0130 \pm 0.0008$ if $R_{2,p} = 1\,\rjup$ ($P_{2,tr} =
0.0122\pm0.0007$ if $R_{2,p} \rightarrow 0$). This derivation assumes
an isotropic inclination distribution. However, dynamical constraints,
such as precise measurements of TTV effects, or analysis of orbital
stability, may further limit the allowed inclination range, and thus
increase or decrease the chance of transits. 

Unfortunately, our HATNet and \flwof\ datasets do not cover any time
interval around the expected transit times of \hatcurc, thus we can not
prove or refute the existence of such transits. Nevertheless, it is an
interesting thought experiment to characterize the putative transits of
\hatcurc, should they occur. \hatcurc\ would be the longest period
transiting planet discovered to date, with a semi-major axis of over
1\,AU, and a period of $\sim \hatcurcLCPshort$\,days, about 4 times
longer than the current record holder \hd{80606}
\citep{naef:2001,fossey:2009,moutou:2009}. With a true mass of
\hatcurcPPmshort\,\mjup, we have good reasons to believe that its
radius would be around 1\rjup, based on heavy-mass transiting planets
like HAT-P-2b \citep[8.84\,\mjup,][]{bakos:2007}, XO-3b
\citep[11.79\,\mjup,][]{johns-krull:2008}, Corot-3b
\citep[21.66\,\mjup,][]{deleuil:2008}, all having radii around
1\,\rjup. If the transits are full (i.e.~not grazing), then the transit
depth would be similar to that of the inner planet \hatcurb. The
duration of the transit could be up to \hatcurcLCdurhrshort\,hours.
Follow-up observations of such transits would require either a
world-wide effort, or uninterrupted space-based observations. The next
transit center, according to the present analysis, will occur at 
2010 April 12 9am UTC. Since we
are looking at the orbit of \hatcurc\ along the semi-latus rectum
(parallel to the minor-axis), the chance for an occultation of the
planet by the star has a very similar chance of occurance as the primary
eclipse. Observations of the secondary eclipse could greatly decrease
the error on $e_2$ and $\omega_2$.

As regards to the nature of \hatcurc, even at its minimal mass
(\hatcurcPPm\,\mjup) it is the 10th most massive planet out of 327
planets listed on the Extrasolar Planet Encyclopedia as of 2009 July.
Curiously, the recently announced Doppler-detection \hd{126614b}
\citep{howard:2009} appears to have similar orbital characteristics to
\hatcurc. Both are Jovian planets in $P > 1$\,yr, $e = 0.7$ orbits
around metal-rich ($\feh\approx 0.5$) stars. As described earlier in
\refsec{ttv}, we have good hopes that in the near future precise TTV
measurements of the inner planet \hatcurb, will constrain the orbital
inclination and thus the true mass of \hatcurc. Further, such TTV
variations can also constrain the mutual inclination of \hatcurb\ and
\hatcurc. Measuring the sky-projected angle between the stellar spin
axis and the orbital normal of \hatcurb\ (the inner planet) via the
Rossiter-McLaughlin effect will shed light on the migration history of
\hatcurb, and by inference the scattering history between \hatcurb\ and
\hatcurc.


\acknowledgements 

HATNet operations have been funded by NASA grants NNG04GN74G,
NNX08AF23G and SAO IR\&D grants. Work of G.\'A.B.~and J.~Johnson were
supported by the Postdoctoral Fellowship of the NSF Astronomy and
Astrophysics Program (AST-0702843 and AST-0702821, respectively). We
acknowledge partial support also from the Kepler Mission under NASA
Cooperative Agreement NCC2-1390 (D.W.L., PI). G.K.~thanks the Hungarian
Scientific Research Foundation (OTKA) for support through grant
K-60750. G.T.\ acknowledges partial support from NASA Origins grant
NNX09AF59G. This research has made use of Keck telescope time granted
through NOAO (program A146Hr,A264Hr) and NASA (N128Hr,N145Hr). We are
grateful to Josh Winn and Matthew Holman for their flexibility in
swapping nights at the \flwof\ telescope.  We thank the anonymous
referee for the useful comments that improved this paper.




\end{document}

%% file: planetinfo.tex
\newcommand{\hatcurfield}{136}                                         
\newcommand{\hatcurCCra}{\ensuremath{08^{\mathrm h}39^{\mathrm m}31.82{\mathrm s}}}                                  
\newcommand{\hatcurCCdec}{\ensuremath{+47{\arcdeg}21{\arcmin}07.4{\arcsec}}}                                 
\newcommand{\hatcurCCtwomass}{2MASS~08393180+4721073}                  
\newcommand{\hatcurCCgsc}{GSC~3416-00543}                              
\newcommand{\hatcurCCtassmv}{10.622}                                   
\newcommand{\hatcurCCtwomassJmag}{\ensuremath{9.328\pm0.018}}          
\newcommand{\hatcurCCtwomassHmag}{\ensuremath{9.058\pm0.017}}          
\newcommand{\hatcurCCtwomassKmag}{\ensuremath{8.975\pm0.017}}          
\newcommand{\hatcurCCcitKmag}{\ensuremath{8.999\pm0.017}}              
\newcommand{\hatcurCCesoJmag}{\ensuremath{9.396\pm0.022}}              
\newcommand{\hatcurCCesoHmag}{\ensuremath{9.070\pm0.021}}              
\newcommand{\hatcurCCesoKmag}{\ensuremath{9.018\pm0.018}}              
\newcommand{\hatcurLCdip}{\ensuremath{6.2}}                            
\newcommand{\hatcurLCrprstar}{\ensuremath{0.0844\pm0.0013}}            
\newcommand{\hatcurLCbsq}{\ensuremath{0.447_{-0.056}^{+0.044}}}        
\newcommand{\hatcurLCimp}{\ensuremath{0.668_{-0.045}^{+0.032}}}        
\newcommand{\hatcurLCzeta}{\ensuremath{17.07\pm0.16}}                  
\newcommand{\hatcurLCdur}{\ensuremath{0.1345\pm0.0017}}                
\newcommand{\hatcurLCdurhr}{\ensuremath{3.228\pm0.040}}                
\newcommand{\hatcurLCq}{\ensuremath{0.0461\pm0.0006}}                  
\newcommand{\hatcurLCingdur}{\ensuremath{0.0180\pm0.0018}}             
\newcommand{\hatcurLCP}{\ensuremath{2.916260\pm0.000010}}              
\newcommand{\hatcurLCPprec}{\ensuremath{2.9162595}}                    
\newcommand{\hatcurLCPshort}{\ensuremath{2.9163}}                      
\newcommand{\hatcurLCT}{\ensuremath{2454779.92979\pm0.00038}}          
\newcommand{\hatcurLCTA}{\ensuremath{2453700.9182\pm0.0068}}           
\newcommand{\hatcurLCTB}{\ensuremath{2453878.8216\pm0.0093}}           
\newcommand{\hatcurLCiblend}{\ensuremath{0.82\pm0.06}}                 
\newcommand{\hatcurLChatnetm}{\ensuremath{9.7966\pm0.0001}}            
\newcommand{\hatcurSMEiteff}{\ensuremath{5760\pm90}}                   
\newcommand{\hatcurSMEizfeh}{\ensuremath{+0.50\pm0.08}}                
\newcommand{\hatcurSMEizfehshort}{\ensuremath{+0.50}}                  
\newcommand{\hatcurSMEilogg}{\ensuremath{4.28\pm0.10}}                 
\newcommand{\hatcurSMEivsin}{\ensuremath{1.9\pm0.5}}                   
\newcommand{\hatcurSMEivmac}{\ensuremath{3.78}}                  
\newcommand{\hatcurSMEivmic}{\ensuremath{0.85}}                  
\newcommand{\hatcurSMEiiteff}{\ensuremath{5653\pm90}}                  
\newcommand{\hatcurSMEiizfeh}{\ensuremath{+0.41\pm0.08}}               
\newcommand{\hatcurSMEiizfehshort}{\ensuremath{+0.41}}                 
\newcommand{\hatcurSMEiilogg}{\ensuremath{4.13\pm0.00}}                
\newcommand{\hatcurSMEiivsin}{\ensuremath{2.9\pm1.0}}                 
\newcommand{\hatcurSMEiivmac}{\ensuremath{3.83}}                 
\newcommand{\hatcurSMEiivmic}{\ensuremath{0.85}}                 
\newcommand{\hatcurDSteff}{\ensuremath{5500\pm100}}                    
\newcommand{\hatcurDSlogg}{\ensuremath{4.0\pm0.25}}                    
\newcommand{\hatcurDSvsini}{\ensuremath{0.5\pm1.0}}                    
\newcommand{\hatcurDSgamma}{\ensuremath{+14.69\pm0.68}}                
\newcommand{\hatcurDSnumspec}{\ensuremath{7}}                          
\newcommand{\hatcurDSspan}{\ensuremath{37}}                            
\newcommand{\hatcurDSrvrms}{\ensuremath{0.68}}                         
\newcommand{\hatcurLBii}{\ensuremath{0.3060}}                          
\newcommand{\hatcurLBiii}{\ensuremath{0.3229}}                         
\newcommand{\hatcurISOm}{\ensuremath{1.22_{-0.10}^{+0.05}}}            
\newcommand{\hatcurISOmlong}{\ensuremath{1.219_{-0.099}^{+0.050}}}     
\newcommand{\hatcurISOr}{\ensuremath{1.56\pm0.08}}                     
\newcommand{\hatcurISOrlong}{\ensuremath{1.559\pm0.082}}               
\newcommand{\hatcurISOlogg}{\ensuremath{4.13\pm0.04}}                  
\newcommand{\hatcurISOlum}{\ensuremath{2.22\pm0.31}}                   
\newcommand{\hatcurISOmv}{\ensuremath{3.97\pm0.17}}                    
\newcommand{\hatcurISOage}{\ensuremath{5.0_{-0.7}^{+2.5}}}             
\newcommand{\hatcurISOMK}{\ensuremath{2.36\pm0.12}}                    
\newcommand{\hatcurISOspec}{G4}                                        
\newcommand{\hatcurRVK}{\ensuremath{106.1\pm1.4}}                      
\newcommand{\hatcurRVk}{\ensuremath{-0.016\pm0.008}}                   
\newcommand{\hatcurRVh}{\ensuremath{0.008\pm0.012}}                    
\newcommand{\hatcurRVgamma}{\ensuremath{-51.3\pm3.8}}                  
\newcommand{\hatcurRVjitter}{\ensuremath{3.0}}                         
\newcommand{\hatcurRVeccen}{\ensuremath{0.021\pm0.009}}                
\newcommand{\hatcurRVomega}{\ensuremath{181\pm45}}                     
\newcommand{\hatcurPPi}{\ensuremath{83.4\pm0.6}}                       
\newcommand{\hatcurPPg}{\ensuremath{12.8_{-1.2}^{+1.6}}}               
\newcommand{\hatcurPPlogg}{\ensuremath{3.11\pm0.05}}                   
\newcommand{\hatcurPPar}{\ensuremath{5.84\pm0.26}}                     
\newcommand{\hatcurPParel}{\ensuremath{0.0427_{-0.0012}^{+0.0006}}}    
\newcommand{\hatcurPPrho}{\ensuremath{0.498_{-0.069}^{+0.103}}}        
\newcommand{\hatcurPPm}{\ensuremath{0.85_{-0.05}^{+0.03}}}             
\newcommand{\hatcurPPmshort}{\ensuremath{0.85}}                        
\newcommand{\hatcurPPmlong}{\ensuremath{0.853_{-0.046}^{+0.029}}}      
\newcommand{\hatcurPPr}{\ensuremath{1.28\pm0.08}}                      
\newcommand{\hatcurPPrshort}{\ensuremath{1.28}}                        
\newcommand{\hatcurPPrlong}{\ensuremath{1.281\pm0.079}}                
\newcommand{\hatcurPPmrcorr}{\ensuremath{0.56}}                        
\newcommand{\hatcurPPteff}{\ensuremath{1653\pm45}}                     
\newcommand{\hatcurPPtheta}{\ensuremath{0.046\pm0.003}}                

\newcommand{\hatcurPPfluxperi}{\ensuremath{(1.76\pm0.20)\cdot10^{13}}}       
\newcommand{\hatcurPPfluxap}{\ensuremath{(1.62\pm0.18)\cdot10^{13}}}         
\newcommand{\hatcurPPfluxavg}{\ensuremath{(1.69\pm0.18)\cdot10^{13}}}        

\newcommand{\hatcurXsecondary}{\ensuremath{2454781.359\pm0.014}}       
\newcommand{\hatcurXsecdur}{\ensuremath{0.1345\pm0.0069}}              
\newcommand{\hatcurXsecingdur}{\ensuremath{0.0180\pm0.0018}}           
\newcommand{\hatcurPPperi}{\ensuremath{2454780.64\pm0.42}}             
\newcommand{\hatcurPPaequiv}{\ensuremath{0.0285\pm0.0016}}             
\newcommand{\hatcurcLCdur}{\ensuremath{0.621\pm0.029}}                 
\newcommand{\hatcurcLCdurhrshort}{\ensuremath{14.9}}                   
\newcommand{\hatcurcLCingdur}{\ensuremath{0.0355\pm0.0012}}            
\newcommand{\hatcurcLCP}{\ensuremath{428.5\pm3.0}}                     
\newcommand{\hatcurcLCPshort}{\ensuremath{428}}                        
\newcommand{\hatcurcLCT}{\ensuremath{2454870.4\pm1.8}}                 
\newcommand{\hatcurcRVK}{\ensuremath{502\pm37}}                        
\newcommand{\hatcurcRVk}{\ensuremath{-0.690\pm0.018}}                  
\newcommand{\hatcurcRVh}{\ensuremath{0.039\pm0.005}}                   
\newcommand{\hatcurcRVeccen}{\ensuremath{0.691\pm0.018}}               
\newcommand{\hatcurcRVomega}{\ensuremath{176.7\pm0.5}}                 
\newcommand{\hatcurcPParel}{\ensuremath{1.188_{-0.033}^{+0.018}}}      
\newcommand{\hatcurcPPm}{\ensuremath{15.2\pm1.0}}                      
\newcommand{\hatcurcPPmshort}{\ensuremath{15.2}}                       
\newcommand{\hatcurcPPmlong}{\ensuremath{15.2\pm1.0}}                  
\newcommand{\hatcurcPPteff}{\ensuremath{340\pm9}}                      

\newcommand{\hatcurcPPfluxperi}{\ensuremath{(2.26\pm0.35)\cdot10^{11}}}      
\newcommand{\hatcurcPPfluxap}{\ensuremath{(7.61\pm0.86)\cdot10^{9}}}        
\newcommand{\hatcurcPPfluxavg}{\ensuremath{(3.00\pm0.33)\cdot10^{10}}}          

\newcommand{\hatcurcXsecondary}{\ensuremath{2454912.7\pm1.8}}          
\newcommand{\hatcurcXsecdur}{\ensuremath{0.574\pm0.028}}               
\newcommand{\hatcurcXsecingdur}{\ensuremath{0.0355\pm0.0012}}          
\newcommand{\hatcurcPPperi}{\ensuremath{2454890.05\pm0.48}}            
\newcommand{\hatcurXdist}{\ensuremath{214\pm12}}                       

%% file: data/rvtable.tex
$ 547.89534 $ \dotfill & $    21.42 $ & $     4.30 $ & $     1.37 $ & $     5.85 $ & $    0.0041 $ & $   0.00005 $\\
$ 548.80173 $ \dotfill & \nodata      & \nodata      & $    -1.90 $ & $     7.27 $ & $    0.0044 $ & $   0.00004 $\\
$ 602.73395 $ \dotfill & $    17.80 $ & $     1.43 $ & $    12.35 $ & $     5.35 $ & $    0.0042 $ & $   0.00002 $\\
$ 602.84690 $ \dotfill & $    22.56 $ & $     1.61 $ & $    -6.03 $ & $     6.73 $ & $    0.0040 $ & $   0.00002 $\\
$ 603.73414 $ \dotfill & $   182.48 $ & $     1.35 $ & $     6.95 $ & $     5.76 $ & $    0.0042 $ & $   0.00002 $\\
$ 603.84323 $ \dotfill & $   202.33 $ & $     2.02 $ & $    -0.77 $ & $     6.52 $ & $    0.0040 $ & $   0.00003 $\\
$ 633.77240 $ \dotfill & $   211.86 $ & $     1.93 $ & $    -4.24 $ & $     7.08 $ & $    0.0039 $ & $   0.00003 $\\
$ 634.75907 $ \dotfill & $    40.20 $ & $     1.95 $ & $    -4.20 $ & $     7.62 $ & $    0.0040 $ & $   0.00003 $\\
$ 635.75475 $ \dotfill & $   183.01 $ & $     2.19 $ & $    -4.47 $ & $     7.63 $ & $    0.0041 $ & $   0.00003 $\\
$ 636.74968 $ \dotfill & $   204.23 $ & $     1.71 $ & $    -2.98 $ & $     6.98 $ & $    0.0041 $ & $   0.00002 $\\
$ 727.13851 $ \dotfill & $   215.53 $ & $     1.81 $ & $    13.22 $ & $     5.05 $ & $    0.0041 $ & $   0.00003 $\\
$ 728.13190 $ \dotfill & $    38.36 $ & $     1.64 $ & $    10.16 $ & $     5.60 $ & $    0.0043 $ & $   0.00003 $\\
$ 778.07302 $ \dotfill & $    42.19 $ & $     1.45 $ & $    -0.61 $ & $     6.46 $ & $    0.0041 $ & $   0.00003 $\\
$ 779.08375 $ \dotfill & $   219.58 $ & $     1.72 $ & $     6.92 $ & $     5.77 $ & $    0.0041 $ & $   0.00003 $\\
$ 780.09369 $ \dotfill & $    87.11 $ & $     1.59 $ & $    11.80 $ & $     5.04 $ & $    0.0042 $ & $   0.00003 $\\
$ 791.11130 $ \dotfill & $   193.44 $ & $     1.66 $ & $    -0.17 $ & $     6.67 $ & $    0.0042 $ & $   0.00002 $\\
$ 809.99576 $ \dotfill & $   -14.92 $ & $     2.33 $ & $     0.27 $ & $     6.54 $ & $    0.0041 $ & $   0.00003 $\\
$ 810.92159 $ \dotfill & $   157.29 $ & $     3.43 $ & $    -2.44 $ & $     6.77 $ & $    0.0046 $ & $   0.00004 $\\
$ 839.06085 $ \dotfill & $  -124.84 $ & $     1.49 $ & $    -8.53 $ & $     7.60 $ & $    0.0041 $ & $   0.00002 $\\
$ 865.02660 $ \dotfill & $  -350.51 $ & $     1.41 $ & $     0.09 $ & $     6.71 $ & $    0.0041 $ & $   0.00002 $\\
$ 867.90311 $ \dotfill & $  -387.07 $ & $     2.55 $ & $   -12.77 $ & $     9.37 $ & $    0.0042 $ & $   0.00003 $\\
$ 928.83635 $ \dotfill & $  -188.44 $ & $     1.35 $ & $     4.00 $ & $     6.19 $ & $    0.0042 $ & $   0.00002 $\\
$ 929.84447 $ \dotfill & $  -189.69 $ & $     2.88 $ & $   -19.31 $ & $     8.75 $ & $    0.0040 $ & $   0.00004 $\\
$ 955.86964 $ \dotfill & $   -90.43 $ & $     1.58 $ & $   -13.92 $ & $     8.19 $ & $    0.0040 $ & $   0.00003 $\\
$ 956.86327 $ \dotfill & $    95.55 $ & $     1.81 $ & $    -5.32 $ & $     7.22 $ & $    0.0039 $ & $   0.00003 $\\
$ 963.85163 $ \dotfill & $   -20.24 $ & $     1.62 $ & $    -5.27 $ & $     7.21 $ & $    0.0041 $ & $   0.00002 $\\
$ 983.74976 $ \dotfill & $   139.67 $ & $     1.47 $ & $     4.02 $ & $     6.21 $ & $    0.0041 $ & $   0.00002 $\\
$ 984.76460 $ \dotfill & $   -35.66 $ & $     1.40 $ & $     6.16 $ & $     5.92 $ & $    0.0040 $ & $   0.00002 $\\
$ 985.73856 $ \dotfill & $   120.74 $ & $     1.39 $ & $     5.20 $ & $     6.05 $ & $    0.0041 $ & $   0.00001 $\\
$ 985.74584 $ \dotfill & \nodata      & \nodata      & $     6.77 $ & $     5.66 $ & $    0.0040 $ & $   0.00002 $\\
$ 985.75333 $ \dotfill & \nodata      & \nodata      & $     6.37 $ & $     5.88 $ & $    0.0042 $ & $   0.00002 $\\
$ 986.76358 $ \dotfill & $   125.20 $ & $     1.73 $ & $     3.95 $ & $     6.31 $ & $    0.0042 $ & $   0.00002 $\\
$ 988.74066 $ \dotfill & $   149.15 $ & $     1.64 $ & $    -7.87 $ & $     7.22 $ & $    0.0041 $ & $   0.00002 $\\

%% file: hatp13.bbl
\begin{thebibliography}{}

\bibitem[Agol et al.(2005)]{agol:2005} Agol, E., Steffen, J., Sari, R., 
\& Clarkson, W.~2005, \mnras, 359, 567

\bibitem[Cox(2000)]{cox:2000} Cox, A.~N.~2000, Allen's Astrophysical Quantities


\bibitem[Bakos et al.(2002)]{bakos:2002}
 Bakos, G.~\'A., L\'az\'ar, J., Papp, I., S\'ari, P.
 \& Green, E.~M.~2002, \pasp, 114, 974

\bibitem[Bakos et al.(2004)]{bakos:2004}
 Bakos, G.~\'A., Noyes, R.~W., Kov\'acs, G., Stanek, K.~Z.,
 Sasselov, D.~D., \& Domsa, I.~2004, \pasp, 116, 266

\bibitem[Bakos et al.(2007)]{bakos:2007} Bakos, G.~{\'A}., et al.~2007, 
\apj, 670, 826 




\bibitem[Bakos et al.(2009)]{bakos:2009} Bakos, G.~{\'A}., et al.~2009,
arXiv:0901.0282



\bibitem[Baraffe et al.(2008)]{baraffe:2008}
Baraffe, I., Chabrier, G., \& Barman, T.~2008, \aap, 482, 315 







\bibitem[Barning(1963)]{barning:1963} Barning, F.~J.~M.~1963, \bain, 17, 22 

\bibitem[Batygin, Bodenheimer, \& Laughlin(2009)]{batygin:2009} Batygin, K., Bodenheimer, P., \& Laughlin, G.~2009, arXiv:0907.5019



\bibitem[Bodenheimer, Lin, \& Mardling(2001)]{bodenheimer:2001} Bodenheimer, P., Lin, D.~N.~C.,
	\& Mardling, R.~A.~2001, \apj, 548, 466 



\bibitem[Borkovits et al.(2003)]{borkovits:2003} Borkovits, T.,
{\'E}rdi, B., Forg{\'a}cs-Dajka, E., \& Kov{\'a}cs, T.~2003, \aap, 398, 1091 










\bibitem[Burke et al.(2007)]{burke:2007} Burke, C.~J., et al.~2007, \apj, 671, 2115





\bibitem[Butler et al.(1996)]{butler:1996} 
Butler, R.~P.~et al.~1996, \pasp, 108, 500



\bibitem[Carpenter(2001)]{carpenter:2001} Carpenter, J.~M.~2001, \aj, 121, 2851 

\bibitem[Carter et al.(2008)]{carter:2008} Carter, J.~A., Yee, J.~C., Eastman, J., Gaudi, B.~S., \& Winn, J.~N.~2008, \apj, 689, 499 








\bibitem[Claret(2004)]{claret:2004}
 Claret, A.~2004, \aap, 428, 1001




\bibitem[Deleuil et al.(2008)]{deleuil:2008} Deleuil, M., Deeg, H.~J., 
Alonso, R., Bouchy, F., \& Rouan, D.~2008, arXiv:0810.0919

\bibitem[Deeming(1975)]{deeming:1975} Deeming, T.~J.~1975, \apss, 36, 137








\bibitem[Droege et al.(2006)]{droege:2006} Droege, T.~F., Richmond,M.~W., 
Sallman, M.~P., \& Creager, R.~P.~2006, \pasp, 118, 1666 



\bibitem[Fabrycky, Johnson, \& Goodman(2007)]{fabrycky:2007} Fabrycky, D.~C., Johnson, E.~T., \& Goodman, J.~2007, \apj, 665, 754 

\bibitem[Fabrycky(2009)]{fabrycky:2009} Fabrycky, D.~C.~2009, IAU Symposium, 253, 173 




\bibitem[Ford(2006)]{ford:2006}
Ford, E.~2006, \apj, 642, 505


\bibitem[Fortney et al.(2007)]{fortney:2007} Fortney, J.~J., Marley, M.~S., \& Barnes, J.~W.~2007, \apj, 659, 1661


\bibitem[Fossey, Waldmann, \& Kipping(2009)]{fossey:2009} Fossey, S.~J., Waldmann, I.~P., \& Kipping, D.~M.~2009, \mnras, 396, L16 









\bibitem[Hansen \& Barman(2007)]{hansen:2007} Hansen, B.~M.~S., \& Barman, T.~2007, \apj, 671, 861 

\bibitem[Hartman et al.(2009)]{hartman:2009} Hartman, J.~D., et al.~2009, 
arXiv:0904.4704 



\bibitem[Holman \& Murray(2005)]{holman:2005} Holman, M.~J., \& Murray, N.~W.~2005, Science, 307, 1288 

\bibitem[Holman et al.(2006)]{holman:2006} Holman, M.~J., et al.\ 2006, \apj, 652, 1715

\bibitem[Howard et al.(2009)]{howard:2009} Howard, W.~A., et al.~2009, 
submitted to \apj



\bibitem[Kane \& von Braun(2009)]{kane:2009} Kane, S.~R., \& von Braun, K.~2009, IAU
Symposium, 253, 358 


\bibitem[Kov\'acs et al.(2002)]{kovacs:2002}
Kov\'acs, G., Zucker, S., \& Mazeh, T.~2002, \aap, 391, 369

\bibitem[Kov\'acs et al.(2005)]{kovacs:2005}
Kov\'acs, G., Bakos, G.~\'A., \& Noyes, R.~W.~2005, \mnras, 356, 557

\bibitem[Johns-Krull et al.(2008)]{johns-krull:2008} Johns-Krull, C.~M., 
et al.~2008, \apj, 677, 657


\bibitem[Latham(1992)]{latham:1992}
 Latham, D.~W.~1992, in IAU Coll.~135, Complementary Approaches to
 Double and Multiple Star Research, ASP Conf.~Ser.~32, 
 eds.~H.~A.~McAlister \& W.~I.~Hartkopf (San Francisco: ASP), 110






\bibitem[McCullough et al.(2006)]{pmcc:2006} McCullough, P.~R.,et al.~2006, \apj, 648, 1228 





\bibitem[P\'al \& Bakos(2006)]{pal:2006}
 P\'al, A., \& Bakos, G.~\'A. 2006, \pasp, 118, 1474



\bibitem[P\'al et al.(2008)]{pal:2008}
P\'al, A., Bakos, G.~\'A., Noyes, R. W. \& Torres, G. 
2008, proceedings of IAU Symp. 253 ``Transiting Planets``, 
ed.~by F. Pont (arXiv:0807.1530)

\bibitem[P{\'a}l(2008)]{pal:2008b} P{\'a}l, A.~2008b, \mnras, 390, 281 


\bibitem[P\'al (2009)]{pal:2009} P\'al, A., PhD thesis,
(arXiv:0906.3486)


\bibitem[Press et al.(1992)]{press:1992}
Press, W. H., Teukolsky, S. A., Vetterling, W. T. \& Flannery, B. P., 1992,
Numerical  Recipes in C: the art of scientific computing,
Second Edition, Cambridge University Press



\bibitem[Mandel \& Agol(2002)]{mandel:2002}
 Mandel, K., \& Agol, E.~2002, \apjl, 580, L171

\bibitem[Mardling(2007)]{mardling:2007} Mardling, R.~A.~2007, \mnras, 382, 1768 





\bibitem[Marcy \& Butler(1992)]{marcy:1992}
 Marcy, G.~W., \& Butler, R.~P.~1992, \pasp, 104, 270







\bibitem[McLaughlin(1924)]{mclaughlin:1924} McLaughlin, D.~B.~1924,
\apj, 60, 22 

\bibitem[Miller, Fortney, \& Jackson(2009)]{miller:2009} Miller, N., Fortney, J.~J., \& Jackson, B.~2009, arXiv:0907.1268


\bibitem[Moutou et al.(2009)]{moutou:2009} Moutou, C., et al.~2009, \aap, 498, L5






\bibitem[Naef et al.(2001)]{naef:2001} Naef, D., et al.~2001, \aap, 375, L27




















\bibitem[Shporer et al.(2008)]{shporer:2008} Shporer, A., et al.~2008, 
arXiv:0806.4008





\bibitem[Skrutskie et al.(2006)]{skrutskie:2006} Skrutskie, M.~F., et 
al.~2006, \aj, 131, 1163


\bibitem[Smith et al.(2009)]{smith:2009} Smith, A.~M.~S., et al.~2009, arXiv:0906.3414





\bibitem[Sozzetti et al.(2007)]{sozzetti:2007}
 Sozzetti, A.~et al.~2007, \apj, 664, 1190

\bibitem[Torres et al.(2002)]{torres:2002}
 Torres, G., Boden, A.~F., Latham, D.~W., Pan, M.~\& Stefanik, R.~P.~2002, \aj, 124, 1716


\bibitem[Torres et al.(2007)]{torres:2007}
 Torres, G.~et al.~2007, \apjl, 666, 121


%


\bibitem[Valenti \& Fischer(2005)]{valenti:2005}
 Valenti, J.~A., \& Fischer, D.~A. 2005, \apjs, 159, 141

\bibitem[Valenti \& Piskunov(1996)]{valenti:1996}
 Valenti, J.~A., \& Piskunov, N.~1996, \aaps, 118, 595

\bibitem[Vogt et al.(1994)]{vogt:1994}
 Vogt, S.~S.~et al.~1994, Proc.~SPIE, 2198, 362

\bibitem[Yi et al.(2001)]{yi:2001}
 Yi, S.~K.~et al.~2001, \apjs, 136, 417








\bibitem[Winn et al.(2007)]{winn:2007} Winn, J.~N., et al.\ 2007, \aj, 133, 1828 

\bibitem[Winn et al.(2009)]{winn:2009} Winn, J.~N., Johnson, J.~A., 
	Albrecht, S., Howard, A.~W., Marcy, G.~W., Crossfield, I.~J., \& 
	Holman, M.~J.~2009, \apjl, 703, L99 






\bibitem[Wittenmyer et al.(2007)]{wittenmyer:2007} Wittenmyer, R.~A., 
Endl, M., Cochran, W.~D., \& Levison, H.~F.~2007, \aj, 134, 1276



\bibitem[Wright et al.(2007)]{wright:2007} Wright, J.~T., et al.~2007, \apj, 657, 533 

\end{thebibliography}
